\documentclass[11pt]{article}
\usepackage{jheppub}
\usepackage{amsmath, amsfonts, amssymb, array}
\usepackage{mathrsfs}
\usepackage{slashed}
\usepackage{multirow}
\newcolumntype{C}[1]{>{\centering\let\newline\\\arraybackslash\hspace{0pt}}m{#1}}
\usepackage[english]{babel}
\usepackage[autostyle]{csquotes}
\usepackage{float}
\usepackage{mathtools}
\usepackage[symbol]{footmisc}
\usepackage{cancel}

\usepackage{hyperref}
\hypersetup{colorlinks=true, linkcolor=blue, urlcolor=blue,citecolor=blue, linktoc=page}

\newcommand{\alphatwo}{{m}}

\newcommand{\be}{\begin{equation}}
\newcommand{\ee}{\end{equation}}
\newcommand{\bea}{\begin{eqnarray}}
\newcommand{\eea}{\end{eqnarray}}
\newcommand{\nn}{\nonumber}

\def\d{\delta}

\def\nn{\nonumber}

\def\nn{\nonumber}

\newcommand{\ben}[1]{\begin{eqnarray}\label{#1} }
\newcommand{\een}{\end{eqnarray}}

\newcommand{\DD}{{\cal D}}

\newcommand{\ga}{\gamma}

\newcommand{\Dslash}{\not{\hbox{\kern-4pt $D$}}}
\newcommand{\pslash}{\not{\hbox{\kern-4pt $\partial$}}}
\newcommand{\Dcslash}{\not{\hbox{\kern-4pt $\DD$}}}

\usepackage{graphicx}
\usepackage{amssymb}
\usepackage{amsthm}
\usepackage{amsmath}
\usepackage{slashed}
\usepackage{color}
\usepackage{tensor}
\usepackage{bbold}
\usepackage[normalem]{ulem}

\newcommand{\phiF}{\phi}

\newcommand{\PhiF}{\Phi}

\newcommand{\varphiF}{\varphi}

\newcommand{\eF}{e}
\newcommand{\eR}{{\boldsymbol{e}}}
\newcommand{\eS}{\hat{\boldsymbol{e}}}

\newcommand{\omegaF}{\omega}
\newcommand{\omegaR}{{\boldsymbol{\omega}}}
\newcommand{\omegaS}{\hat{\boldsymbol{\omega}}}

\newcommand{\bbF}{b}
\newcommand{\bbR}{{\boldsymbol{b}}}

\newcommand{\VF}{V}
\newcommand{\VR}{{\boldsymbol{V}}}
\newcommand{\VS}{\widehat{\boldsymbol{V}}}

\newcommand{\ffF}{f}

\newcommand{\aaF}{a}

\newcommand{\EF}{E}
\newcommand{\ER}{{\boldsymbol{E}}}

\newcommand{\TF}{T}
\newcommand{\TR}{{\boldsymbol{T}}}
\newcommand{\TS}{\widehat{\boldsymbol{T}}}

\newcommand{\DF}{D}
\newcommand{\DR}{{\boldsymbol{D}}}
\newcommand{\DS}{\widehat{\boldsymbol{D}}}

\newcommand{\betaR}{{\boldsymbol{\beta}}}
\newcommand{\betaS}{\widehat{\beta}}

\newcommand{\epsR}{{\boldsymbol{\epsilon}}}
\newcommand{\epsS}{\hat{\boldsymbol{\epsilon}}}

\newcommand{\psiF}{\psi}
\newcommand{\psiR}{{\boldsymbol{\psi}}}
\newcommand{\psiS}{\widehat{\boldsymbol{\psi}}}

\newcommand{\LambdaF}{\Lambda}

\newcommand{\chiF}{\chi}
\newcommand{\chiR}{{\boldsymbol{\chi}}}
\newcommand{\chiS}{\widehat{\boldsymbol{\chi}}}

\newcommand{\PF}{P}

\newcommand{\AF}{A}
\newcommand{\AR}{{\boldsymbol{A}}}

\newcommand{\GF}{G}

\newcommand{\FFF}{F}

\newcommand{\rhoF}{\rho}

\newcommand{\epsilonF}{\epsilon}

\newcommand{\epsilonS}{\widehat{\boldsymbol{\epsilon}}}

\newcommand{\XF}{X}

\newcommand{\etaF}{\eta}
\newcommand{\etaR}{{\boldsymbol{\eta}}}
\newcommand{\etaS}{\widehat{\boldsymbol{\eta}}}

\newcommand{\sR}{{\boldsymbol{s}}}
\newcommand{\LR}{{\boldsymbol{L}}}

\newcommand{\sigmaR}{{\boldsymbol{\sigma}}}
\newcommand{\tauR}{{\boldsymbol{\tau}}}

\newcommand{\Cayley}{\mathcal{C}}

\numberwithin{equation}{section}

\def\d{\delta}

\makeatletter
\g@addto@macro\bfseries{\boldmath}
\makeatother

\title{Torus reduction of maximal conformal supergravity}

\author[1]{Franz Ciceri,}
\author[2,3]{Axel Kleinschmidt,}
\author[4,*]{Subrabalan Murugesan\note{Present address: Heriot-Watt University, Edinburgh, Scotland, UK EH14 4AS}}
\author[4]{and Bindusar Sahoo}

\affiliation[1]{ENSL, Laboratoire de physique, F-69342 Lyon, France}
\affiliation[2]{Max Planck Institute for Gravitational Physics (Albert Einstein Institute),
Potsdam Science Park, Am M\"uhlenberg 1, 14476, Potsdam, Germany}
\affiliation[3]{ International Solvay Institutes,
ULB-Campus Plaine CP231, 1050 Brussels, Belgium}
\affiliation[4]{
School of Physics, Indian Institute of Science Education and Research Thiruvananthapuram,
Vithura, Thiruvananthapuram - 695551, India}

\emailAdd{franz.ciceri@ens-lyon.fr}
\emailAdd{axel.kleinschmidt@aei.mpg.de}
\emailAdd{sm2160@hw.ac.uk}
\emailAdd{bsahoo@iisertvm.ac.in}

\abstract{We consider the dimensional reduction of $N=(2,0)$ conformal supergravity in six dimensions on a two-torus to $N=4$ conformal supergravity in four dimensions. At the level of kinematics, the six-dimensional Weyl multiplet is shown to reduce to a mixture of the $N=4$ Weyl and vector multiplets, which can be reinterpreted as a new off-shell multiplet of $N=4$ conformal supergravity. Similar multiplets have been constructed in other settings and are referred to as dilaton Weyl multiplets. We derive it here for the first time in a maximally supersymmetric context in four dimensions. Furthermore, we present the non-linear relations between all the six- and four-dimensional bosonic and fermionic fields, that are obtained by comparing the off-shell supersymmetry transformation rules.}

\begin{document}
	\allowdisplaybreaks
	\maketitle

\renewcommand{\thefootnote}{\arabic{footnote}}	
	
	\section{Introduction}
Conformal supergravities are extensions of Poincaré supergravities that are invariant under local Weyl transformations 
and are constructed by gauging the superconformal algebras~\cite{PhysRevD.17.3179,deWit:1979dzm,Freedman:2012zz}. The associated 
 gauge fields, which include the vielbein and gravitini, are packaged together with auxiliary fields into the so-called Weyl multiplet. The set of local superconformal transformations closes off-shell on the fields of the Weyl multiplet, that is without the need to impose equations of motion. In other words closure\ of supersymmetry is ensured independently of the dynamics considered, and as a result conformal supergravities provide a powerful framework for constructing higher-derivative supersymmetric invariants.

This off-shell formalism can be leveraged to study the higher-derivative structure of Poincaré supergravities\footnote{For a recent review of higher-derivative supergravities see~\cite{Ozkan:2024euj}.} efficiently. This is achieved by considering conformal supergravities coupled to specific systems of matter multiplets. The latter supply fields that act as compensators for the additional local Weyl transformations, and in this way ensure the gauge-equivalence with the Poincaré theories \cite{Kaku:1978ea,Freedman:2012zz}. The standard Poincaré formulations are recovered by imposing gauge-fixing conditions on the compensators, and subsequently integrating out the auxiliary fields. 
This off-shell approach to the construction of higher-derivative Poincaré invariants played an important role in the study of subleading corrections to the entropy of various BPS black holes. It was for instance applied in the context of asymptotically AdS black holes in $N=2$ (gauged) supergravities in $D=4$ and 5 spacetime dimensions (see for instance \cite{Cremonini:2008tw,Bobev:2021oku,Bobev:2021qxx,Liu:2022sew,Cassani:2022lrk,Gold:2023ymc,Ma:2024ynp}). In order to apply such techniques to broader classes of examples, it is desirable to gather an exhaustive understanding of the multiplet structures of conformal supergravities in various dimensions.

In this paper, we focus on $N=4$ conformal supergravity in four dimensions \cite{Bergshoeff:1980is,Buchbinder:2012uh,Ciceri:2015qpa,Butter:2016mtk,Butter:2019edc} and $N=(2,0)$ conformal supergravity in six dimensions \cite{Bergshoeff:1999db,Butter:2017jqu}. They are maximally supersymmetric theories, whose Weyl multiplets both carry 128+128 off-shell degrees of freedom.\footnote{The off-shell counting corresponds to the number of field components minus the number of gauge transformations.} The $D=4$, $N=4$ multiplet however stands apart from all other maximal Weyl multiplets due to the fact that it contains scalar fields which parametrize an $SU(1,1)/U(1)$ coset space, and which are inert under local Weyl transformations. This naturally suggests a relation between the dimensional reduction of $N=(2,0)$ conformal supergravity on $T^2$ and $N=4$ conformal supergravity in four dimensions. We will make this relation explicit at the kinematical level, and as a by product, construct the so-called dilaton Weyl multiplet of $D=4$, $N=4$  conformal supergravity.
	
Dimensional reductions of conformal supergravities were previously considered in the context of half-maximal\footnote{We are talking here about conformal supergravities, which possess standard $Q$-supercharges, as well as special $S$-supercharges. By half-maximal theories, we mean those with eight $Q$-supercharges.} theories \cite{Banerjee:2011ts}. There, the $N=1$ conformal supergravity in five dimensions was shown to reduce on the circle to $N=2$ conformal supergravity in four dimensions.
In this case, the number of off-shell degrees of freedom in the five-dimensional $N=1$ Weyl multiplet is $32+32$, and exceeds that of the four-dimensional $N=2$ Weyl multiplet which only contains $24+24$. Upon reduction, the five dimensional multiplet was shown to decompose into the $N=2$ Weyl multiplet as well as an off-shell $N=2$ vector multiplet in four dimensions that provides the remaining 8+8 degrees of freedom. In particular the four-dimensional vector field and dilaton, which follow from the standard Kaluza--Klein decomposition of the fünfbein, sit inside the vector multiplet. 

A subtle difference arises when considering the reduction of maximally supersymmetric theories. In our case, the degrees of freedom in the $N=(2,0)$ Weyl multiplet in six dimensions already match those of the $N=4$ Weyl multiplet in four dimensions. At first sight, one might then expect these two mutiplets to be directly related by a reduction on $T^2$, just as for maximally supersymmetric Poincaré theories. However this cannot be the case as the four-dimensional $N=4$ Weyl multiplet does not contain vector fields, nor does it contain a scalar field that could serve as the dilaton. In other words, there is no scalar which is a R-symmetry singlet with a non-vanishing Weyl weight. In fact, it is also unclear at this point how the $SU(4)$ R-symmetry group of the four-dimensional theory emerges from the reduction of the six-dimensional theory whose R-symmetry group is $USp(4)$. 

It turns out that as in the half-maximal case, the dimensional reduction of the $N=(2,0)$ Weyl multiplet can also be seen to lead to an $N=4$ Weyl multiplet and a vector multiplet in four dimensions. The crucial difference with the half-maximal case is that the $N=4$ vector multiplet is on-shell, meaning that its fields satisfy certain two-derivative equations of motion involving the Weyl fields as background fields \cite{deRoo:1984zyh}. As a result, it is no longer possible to reason based on an off-shell counting of degrees of freedom carried by the sum of the two multiplets. We will show however that by interpreting the vector multiplet field equations as constraints for some of the auxiliary Weyl multiplet fields, the remaining set of independent fields recombine into a single new $N=4$ off-shell multiplet which indeed carries 128+128 degrees of freedom. In particular, the vector field and its dual are promoted to off-shell fields in the process. An important point is that, in order to carry out this procedure, it is first necessary to impose the following gauge-fixing condition on the $N=4$ vector multiplet scalar $\phi^{ij}$ that transforms in the \textbf{6} of $SU(4)$,
\begin{align}
\phi^{ij}=\frac14 \Omega^{ij}\rho\,,\label{intro:gf}
\end{align}
where $\Omega^{[ij]}=\Omega^{ij}$ denotes the $Sp(4,\mathbb C)$ invariant tensor. The condition \eqref{intro:gf} breaks the $SU(4)$ R-symmetry group to $USp(4)$ and the scalar $\rho$, which carries a non-vanishing Weyl weight, is identified with the dilaton. Together with the vectors fields, it becomes part of the new $N=4$ off-shell multiplet which is therefore referred to as the $N=4$ dilaton Weyl multiplet. The latter makes direct contact with the $T^2$ reduction of the six-dimensional $N=(2,0)$ Weyl multiplet. One of the main results of this paper is to establish the complete non-linear dictionary between the $D=6$ $(2,0)$ Weyl and the $D=4$, $N=4$ dilaton Weyl multiplet, based on the comparison of their superconformal transformations rules.

This work presents the first example of a maximally supersymmetric dilaton Weyl multiplet. Its construction essentially follows the same logic as the one used to build dilaton Weyl multiplets in the context of half-maximally supersymmetric theories. The main difference here is the need to partially break the $SU(4)$ R-symmetry group, by imposing the condition \eqref{intro:gf} which has no analogue in the half-maximal constructions. The latter were carried out in $D=6$, $N=(1,0)$ \cite{BERGSHOEFF1986653}, $D=5$, $N=1$ \cite{Bergshoeff:2001hc}, and $D=4$, $N=2$ \cite{Butter:2017pbp} conformal supergravities. Other examples known as hyperdilaton Weyl multiplets, which were engineered using on-shell hypermultiplets instead of vectors, have also been derived in $D=4$ \cite{Gold:2022bdk}, and $D=5,6$ \cite{Hutomo:2022hdi} conformal supergravities. 

In principle, dilaton Weyl multiplets can allow for new off-shell descriptions of (higher-derivative) supersymmetric invariants. In $D=4$ $N=4$ conformal supergravity, the most general class of off-shell invariants that can be constructed out of the standard Weyl multiplet fields was derived in \cite{Butter:2016mtk,Butter:2019edc}. These four-derivative invariants are characterized by an arbitrary holomorphic function of the coset scalar fields, which in particular appears in front of the leading Weyl tensor squared term. It would be interesting to understand if there exists other $N=4$ off-shell invariants, which could be constructed using the dilaton Weyl multiplet.
 
The plan of the paper is as follows. In section~\ref{sec:construction}, we first review the field content of the $N=4$ Weyl and vector multiplet in four dimensions. We then discuss the R-symmetry gauge-fixing and subsequently construct the dilaton Weyl multiplet. Section~\ref{sec:dw6d} presents the reduction of the six dimensional $N=(2,0)$ Weyl mutiplet on $T^2$. We then derive the non-linear dictionary that relates its fields to those of the four-dimensional dilaton Weyl multiplet. We conclude with a short discussion of potential future applications.

\section{\texorpdfstring{$N=4$ conformal supergravity multiplets in four dimensions}{N=4 conformal supergravity multiplets in four dimensions}}
\label{sec:construction}

$N=4$ conformal supergravity in four spacetime dimensions is built upon the gauging of the $\mathfrak{su}(2,2|4)$ superconformal algebra. Its bosonic subalgebra contains the generators of the conformal group $SU(2,2)$ and of a chiral $SU(4)$ R-symmetry. The fermionic generators consist of sixteen  ordinary $Q$-supercharges and sixteen special $S$-supercharges that appear as the components of two sets of Majorana spinors  whose chiral projections transform in the (anti)fundamental representation of $SU(4)$, labelled here with $i=1,\ldots,4$. The chirality gamma matrix in $D=4$ is denoted by $\gamma_*$. 

\subsection{\texorpdfstring{$N=4$ Weyl multiplet}{N=4 Weyl multiplet}}
    \label{sec:Weyl}
The gauge fields associated to the local $N=4$ superconformal symmetries form, together with various auxiliary fields, the so-called $N=4$ Weyl multiplet. It is an off-shell multiplet, whose construction was described in detail in \cite{Bergshoeff:1980is}. Here, we only summarize the properties of its various fields in Table~\ref{table:gaugefields}, and refer to \cite{Ciceri:2015qpa} for their supersymmetry transformations.
\vspace*{3mm}
	\begin{table}[ht!]
 \begin{center}
   \scalebox{0.80}{
 \begin{tabular}{ c | c c l c c c }
 \hline 
 \hline
   & Field & Gauge symmetry & Name/Restrictions & $SU(4)$ & $w$ & $\mathrm{c}$ \\
 \hline
 \multirow{14}{*}{Bosons} & $\eF_\mu{}^a$ & Translations & vierbein & $\mathbf{1}$ & $-1$ & 0 \\
 & $\omegaF_{\mu}{}^{ab}$  & local Lorentz & spin connection & $\mathbf{1}$ & 0 & 0 \\
 & $\bbF_{\mu}$ & Dilatations & dilatational gauge field & $\mathbf{1}$ & 0 & 0 \\
 & $\VF_{\mu}{}^i{}_j$ & SU(4) & $\mathrm{SU}(4)$ gauge field; $\VF_\mu{}^i{}_i=0$  & $\mathbf{15}$ & 0 & 0 \\
 & & &  $\VF_{\mu}{}_i{}^j\equiv(\VF_{\mu}{}^i{}_j)^\ast=-\VF_\mu{}^j{}_i$  & & & \\
 & $\ffF_{\mu}{}^a$ & conformal boosts & K-gauge field & $\mathbf{1}$ & 1 & 0 \\
 & $\aaF_\mu$ & U(1)& $\mathrm{U}(1)$ gauge field & $\mathbf{1}$ & 0 & 0 \\
 & $\phiF_{\alpha}$ & & $\phiF_\alpha\,\phiF^{\alpha}=1\,,\,\phiF^1=\phiF_1^*\,,\phiF^2=-\phiF_2^*$ & $\mathbf{1}$ & 0 & $-1$ \\
 & $\EF_{ij}$ & & $\EF_{ij}=\EF_{ji}$ & $\overline{\mathbf{10}}$ & 1 & $-1$ \\
 &$\TF_{ab}{}^{ij}$ & & $\tfrac12\varepsilon_{abcd}\TF^{cd}{}^{ij}=-\TF_{ab}{}^{ij}$ & $\mathbf{6}$ & 1 & $-1$\\
 & & & $\TF_{ab}{}^{ij}=-\TF_{ab}{}^{ji}$ & & & \\
 & $\DF^{ij}{}_{kl}$ & & $\DF^{ij}{}_{kl}=\tfrac14 \varepsilon^{ijmn}\varepsilon_{klpq}\DF^{pq}{}_{mn}$ & $\mathbf{20}^\prime$ & 2 & $0$ \\
 & & & $(\DF^{kl}{}_{ij})^\ast=\DF^{ij}{}_{kl}$ & & & \\
 & & & $\DF^{ij}{}_{kj}=0$ & & & \\[1mm]
 \hline
\multirow{4}{*}{Fermions} 
 & $\phiF_{\mu\,i}$ & $S$-supersymmetry & $S$-gauge field; $\ga_*\,\phiF_{\mu i}=\phiF_{\mu `i}$ & $\overline{\mathbf{4}}$ & $\tfrac12$ & $\frac{1}{2}$ \\
& $\psiF_{\mu}{}^i$ & $Q$-supersymmetry & gravitini; $\ga_*\,\psiF_{\mu}{}^i=\psiF_\mu{}^i$ & $\mathbf{4}$ & $-\frac12$ & $-\frac12$ \\
& $\LambdaF_i$  & & $\ga_*\LambdaF_i=\LambdaF_i$ & $\overline{\mathbf{4}}$ & $\tfrac{1}{2}$ & $-\tfrac32$ \\
& $\chiF_k{}^{ij}$  & & $\ga_*\chiF_k{}^{ij}=\chiF_k{}^{ij}$; $\chiF_k{}^{ij}=-\chiF_k{}^{ji}$ & $\mathbf{20}$ & $\tfrac32$ & $-\tfrac12$ \\[1mm]
& & & $\chiF_j{}^{ij}=0$ & & & \\
 \hline
 \hline
\end{tabular}}
$ $\newline
\caption{\textit{Fields of the $N=4$ Weyl multiplet. The last columns denote the Weyl weight $w$ and the chiral weight $\rm c$. The former is the eigenvalue under the dilatation operator of the conformal symmetry while the latter is the eigenvalue under the chiral $U(1)$ appearing in~\eqref{Omegadef}.}}
\label{table:gaugefields}
\end{center}
\end{table}

Among the auxiliary fields of the Weyl multiplet is an $SU(1,1)$ doublet of complex scalars $\phi_\alpha$, with $\alpha=1,2$, which plays a distinguished role for the dynamics of the theory (see for instance \cite{Butter:2016mtk}). These scalars have a vanishing Weyl weight $w=0$, and satisfy the $SU(1,1)$ invariant constraint 
\begin{align}
\phi_\alpha\phi^\alpha=1\,,\;\;\;\;\text{where}\;\phi^\alpha=\eta^{\alpha\beta}(\phi_{\beta})^*\,,\label{SUcons}
\end{align}
where $\eta_{\alpha\beta}=\text{diag}(1,-1)$ is the $SU(1,1)$ invariant  metric. Importantly, these scalars are also subject to an additional local chiral $U(1)$ transformation. Their chiral weights are given in Table~\ref{table:gaugefields}. This implies that the following $SU(1,1)$ element
\begin{align}
	    U=\begin{pmatrix}  \phiF_{1} & -\phiF^{2} \\ \phiF_{2} &   \phiF^{1} \end{pmatrix}\label{Udef1} \,.
\end{align}
transforms under $g\in SU(1,1)$ and local $U(1)$ as
\begin{align}
	    U(x)\to g\,U(x)\,\Omega(x)\;, \label{coset_transf}  \,
\end{align}
with
\begin{align}
	    \Omega(x)=\begin{pmatrix} e^{-i\beta(x)} & 0\\ 0 & e^{i\beta(x)}\end{pmatrix}\,.
	    \label{Omegadef}
	\end{align}
The scalars $\phi_\alpha$ can therefore be seen as coordinates on the coset space $SU(1,1)/U(1)$. We will therefore refer to them as coset scalars. Note that they are the only fields of the Weyl multiplet that transform under $SU(1,1)$. For later purposes, we also define the following combinations 
	\begin{align}
	    \PhiF\coloneqq\phiF^{1}+\phiF^2\;,\; \;\;\;\varphiF\coloneqq\phiF^1-\phiF^2\;.
	    \label{Phidef}
	\end{align}

As in the pure conformal gravity case (see for instance~\cite{Mohaupt:2000mj}), not all of the gauge fields appearing in Table~\ref{table:gaugefields} are independent fields. In particular, the gauge fields corresponding to Lorentz transformations, conformal boosts and $S$-supersymmetry are expressed in terms of the others through a set of conventional supercovariant constraints involving the various curvatures \cite{Bergshoeff:1980is,Ciceri:2015qpa,Butter:2019edc}. The gauge field $a_\mu$ associated to the $U(1)$ factor of $SU(1,1)/U(1)$ is also a dependent field, as is usual in coset space constructions. In the supersymmetry transformation rules, the scalars $\phi_\alpha$ only appear through the $SU(1,1)$ invariant (projection of the) Maurer--Cartan form $\PF_{\mu}$ and its complex conjugate $(P_\mu)^*\equiv\bar{\PF}_{\mu}$ that describe the physical sector of the coset space. They read
\begin{align}
    \PF_{\mu}=\varepsilon_{\alpha\beta}\phiF^{\alpha}D_{\mu}\phiF^{\beta}\;, \;\;\; \bar{\PF}_{\mu}=-\varepsilon^{\alpha\beta}\phiF_{\alpha}D_{\mu}\phiF_{\beta}\,,    \label{Pdef}
\end{align}
where $D_{\mu}$ denotes the fully supercovariant derivative which generally includes the connections for local Lorentz, R-symmetry, dilatation, special conformal transformations, and $Q$- and $S$-supersymmetry as well as the local $U(1)$ symmetry. The $SU(1,1)$ invariant Levi--Civita tensor is defined as $\varepsilon^{12}=\varepsilon_{12}=1$. The forms \eqref{Pdef} satisfy the superconformal generalization of the Maurer--Cartan equation 
\begin{align}
D_{[\mu}P_{\nu]}=\frac12 \bar\Lambda^i\gamma_{[\mu}\Lambda_i \,P_{\nu]}+\bar\Lambda^i R(Q)_{\mu\nu\,i}\,,
\end{align}
and its complex conjugate. The expression of the supercovariant curvature tensor $R(Q)_{\mu\nu\,i}$ associated to $Q$-supersymmetry can be found in~\cite{Ciceri:2015qpa}. The bar on fermions generally denotes the Majorana conjugate. It is defined as
\begin{align}
\bar{\Lambda}_i \coloneqq \Lambda_i^T c\,,
\end{align}
where $c$ is the (real antisymmetric) charge conjugation matrix. Note also that Table~\ref{table:gaugefields} only displays the fermions of positive chirality. They are related to those of negative chirality through charge conjugation,
\begin{align}
\Lambda^i = (\Lambda_i)^C = i \gamma^0 c^{-1} (\Lambda_i)^*\,.
\end{align}
Throughout the paper we will use a (chiral) notation where in four dimensions complex conjugation flips the position of the $SU(4)$ indices, turning for instance the ${\bf 4}$ of $SU(4)$ into the conjugate $\bar{\bf 4}$.

As mentioned in the introduction, what will be relevant in the relation to six-dimensional $N=(2,0)$ conformal supergravity is not the standard four-dimensional $N=4$ Weyl multiplet presented here, but a variant that is usually referred to as the dilaton Weyl multiplet. In order to construct the latter, we will take a cue from the construction of such multiplets in less supersymmetric cases in six, five and four dimensions~\cite{BERGSHOEFF1986653,Bergshoeff:2001hc,Butter:2017pbp}. There, it is obtained by coupling a vector multiplet to conformal supergravity, and by subsequently using the equations of motion of the vector multiplet to trade some of the auxiliary fields of the Weyl multiplet for fields of the vector multiplet. In particular the dual of the vector field gets promoted to an independent field in the process, and becomes part of the off-shell dilaton Weyl multiplet. The procedure for the four-dimensional $N=4$ case will be carried out explicitly in section~\ref{sec:dilWeylmu}.    
	
\subsection{\texorpdfstring{$N=4$ vector multiplet}{N=4 vector multiplet}}
\label{sec:vmultiplet}

The $N=4$ vector multiplet in four-dimensional flat space was originally constructed in \cite{Brink:1976bc}. When coupled to $N=4$ conformal supergravity, its various fields transform under the local superconformal symmetries described in the beginning of the section. We list the vector multiplet fields, together with their weights under dilatation and chiral $U(1)$ transformation in Table~\ref{table:vmfields}. Their supersymmetry transformations, which involve the fields of the $N=4$ Weyl multiplet, are given in \cite{deRoo:1984zyh}.	\vspace*{3mm}
\begin{table}[ht!]
 \begin{center}
 \begin{tabular}{ c c c c c c }
 \hline 
 \hline
 Field & Type & Properties & $SU(4)$ & $w$ & $\mathrm{c}$ \\
 \hline
 $\AF_\mu$ & Boson & Vector gauge field & $\mathbf{1}$ & 0 & 0\\
 $\psiF_i$ & Fermion & $\gamma_*\psiF_i=-\psiF_i$ & $\mathbf{\bar 4}$ & $\frac32$ & $-\frac{1}{2}$ \\
 $\phiF_{ij}$ & Boson & $\phiF^{ij}=(\phiF_{ij})^*=-\tfrac12\,\varepsilon^{ijkl}\phiF_{kl}$ & $\mathbf{6}$ & 1 & 0 \\[1mm]
 \hline
 \hline
\end{tabular}
\center{\caption{\textit{Fields of the $N=4$ vector multiplet.}}
\label{table:vmfields}}
\end{center}
\end{table}

\vspace{-0.3cm}
The vector multiplet is an on-shell multiplet, \textit{i.e.} the superconformal algebra only closes on its fields modulo their equations of motion. In the following we denote the associated two-derivative Lagrangian density, which was constructed in \cite{deRoo:1984zyh}, by $\mathcal{L}$. This Lagrangian also involves the Weyl mutiplet fields which appear as non-propagating background fields. Here, we will only recall the field equations of the vector multiplet, as they directly play a role in the construction of the $N=4$ dilaton Weyl multiplet. The equation of motion and the Bianchi identity for the vector field $A_\mu$ can respectively be written as
\begin{align}
		D_a (  \GF^{+ ab} -  \GF^{- ab}) &= 0\,,\label{eq:eomvec} \\
		D_a (  \FFF^{+ ab} -  \FFF^{- ab}) &= 0 	\,,\label{eq:BI}		
\end{align}
in terms of the supercovariant field strength $\FFF_{\mu\nu}$ and its dual $\GF_{\mu\nu}$. In tangent space, the former reads
\begin{equation}
 \FFF_{ab}=2\,\eF_{[a}{}^{\mu}\eF_{b]}{}^{\nu}\partial_{\mu}\AF_{\nu}-\left[\PhiF\left(\bar{\psiF}_{[a}^{i}\gamma_{b]}\psiF_{i}-\bar{\psiF}_{a}^{i}\psiF_{b}^{j}\phiF_{ij}+\bar{\psiF}_{i[a}\gamma_{b]}\Lambda_{j}\phiF^{ij}\right)+\text{h.c}\right]\,, \label{eq:Fdef}
\end{equation}
while the dual field strength is defined through the Lagrangian $\mathcal L$ as
\begin{align}
	    \GF_{ab}\coloneqq\frac{i}{e}\varepsilon_{abcd}\frac{\delta \mathcal{L}}{\delta \FFF^{}_{cd}}\,,\label{Gdef}
\end{align}
where $e$ denotes the determinant of the vierbein. We use $\varepsilon_{abcd}=i\epsilon_{abcd}$, where $\epsilon_{abcd}$ is the completely antisymmetric standard Levi--Civita symbol in four dimensions with signature $(-+++)$ that satisfies $\epsilon_{0123}=1$. The superscripts $+(-)$ appearing in~\eqref{eq:eomvec} and~\eqref{eq:BI} denote the self-dual (anti-selfdual) projection of the field strengths. For an arbitrary antisymmetric tensor $t_{ab}$,
these projections are defined as
	\begin{align}
	    & t^{\pm}_{ab}\equiv\frac{1}{2}\left(t_{ab}\pm \frac12\varepsilon_{abcd}\,t^{cd}\right)\,.
	    \end{align} 
Note that if $t_{ab}$ is real, complex conjugation implies $(t_{ab}^\pm)^*=t_{ab}^\mp$. Hence, the field equation~\eqref{eq:eomvec} and the Bianchi identity~\eqref{eq:BI} are purely imaginary. From the expression of the Lagrangian given in \cite{deRoo:1984zyh}, it follows that
\begin{align}
 \GF^{+}_{ab}=\frac{2i}{e}\frac{\delta \mathcal{L}}{\delta \FFF^{+}_{ab}}=- i\frac{\varphi}{\PhiF}\FFF^{+}_{ab}+\frac{i}{2\Phi}\bar{\LambdaF}^{i}\gamma_{ab}\psiF_{i}-\frac{2i}{\Phi}\TF_{abij}\phiF^{ij}\,,\label{eq:G+}
	\end{align}
where the scalar combinations $\Phi$ and $\varphi$ were defined in~\eqref{Phidef}. The expression of $\GF^{-}_{ab}$ directly follows from complex conjugation of~\eqref{eq:G+}. The rigid $SU(1,1)$ transformations that we discussed earlier leave the combined set of equations \eqref{eq:eomvec} and \eqref{eq:BI} invariant. Their action on the field strength and its dual is more conveniently described in terms the $SU(1,1)$ doublet $\mathcal{F}_{ab;\alpha}$, whose components read
	\begin{align}\label{F12def}
	    \mathcal{F}_{ab;1}\coloneqq \frac{1}{2}\left(i\GF_{ab}-\FFF_{ab}\right)\;, \quad \mathcal{F}_{ab;2}\coloneqq \frac{1}{2}\left(i\GF_{ab}+\FFF_{ab}\right)\,.
	\end{align}
The transformation reads
	\begin{align}
	   \begin{pmatrix} \mathcal{F}_{ab;1} \\\mathcal{F}_{ab;2}\end{pmatrix}\to g \begin{pmatrix} \mathcal{F}_{ab;1}\\\mathcal{F}_{ab;2}\end{pmatrix}\,,\label{FieldSUtransf}
	\end{align}
where the action of $g\in SU(1,1)$ on the coset scalars is given by \eqref{coset_transf}.
As a result, the following combination of coset scalars and field strengths is manifestly $SU(1,1)$ invariant
	\begin{align}\label{FFF}
    \widetilde{\FFF}_{ab} \coloneqq \varepsilon^{\alpha\beta} \phi_\alpha \mathcal{F}_{ab;\beta}\,.
	\end{align} 
 The complex conjugate reads $(\widetilde F_{ab})^*=\phi^\alpha \mathcal{F}_{ab;\alpha}$.

Let us now move to the equations of motion for the gaugino $\psi_{i}$ and the scalar field $\phi_{ij}$. For the gaugino we have
	\begin{align}
		&\slashed{D}\psiF_i + \frac{1}{4}\gamma\cdot\widetilde{\FFF}\, \Lambda_i + \frac{1}{2}\EF_{ij}\psiF^j -\frac{1}{4}\varepsilon_{ijkl}\gamma\cdot \TF^{jk}\psiF^l \nonumber\\
		&\quad - \chiF\indices{_i^{kl}}\phiF_{kl} + \frac{1}{6}\phiF_{ij}\EF^{jk}\LambdaF_k +\frac{1}{3} \slashed{\bar{\PF}}\phiF_{ij}\LambdaF^j - \frac{1}{8}\gamma_a \psiF_j \bar\LambdaF_i\gamma^a\LambdaF^j = 0 \,,\label{eq:eomchi}
	\end{align}
where for any vector $v_a$, we use the notation $\slashed{v}=\gamma^av_a$ and for any antisymmetric tensor $t_{ab}$, we use $\gamma\cdot t=\gamma^{ab}t_{ab}$.
Because $\Lambda_i$ has positive chirality, the above equation only contains the projection to the anti-selfdual part $\widetilde{F}^-$ of the $SU(1,1)$ invariant (\ref{FFF}). 
The scalar field equation reads
	\begin{align}
		&\Box\phiF_{ij} - (\TF_{ij}\cdot \widetilde{\FFF} - \frac{1}{2}\varepsilon_{ijkl}\TF^{kl}\cdot \widetilde{\FFF}) + (\bar{\chiF}\indices{^k_{ij}}\psiF_k - \frac{1}{2}\varepsilon_{ijkl}\bar\chiF\indices{_m^{kl}}\psiF^m) \nonumber\\
		&\quad+ \frac{1}{2}\DF\indices{^{kl}_{ij}}\phiF_{kl} - \frac{1}{6}(2\bar \psiF_{[i} \slashed{\PF} \LambdaF_{j]} - \varepsilon_{ijkl} \bar \psiF^{k} \slashed{\bar{\PF}} \LambdaF^l) - \frac{1}{12}(2\bar \LambdaF^k\psiF_{[i}\EF_{j]k} - \varepsilon_{ijkl}\bar\LambdaF_m\psiF^k \EF^{lm})\nonumber\\
		&\quad- \frac{1}{12}\phiF_{ij}\EF_{kl}\EF^{kl} + \frac{1}{3}\phiF_{ij}\PF_{a}\bar{\PF}^{a} + \frac{1}{12}\phiF_{ij}(\bar \LambdaF^k\slashed{D} \LambdaF_k + \bar{\LambdaF}_{k}\slashed{D}\LambdaF^{k}) + \frac{1}{8}\phiF_{ij} \bar{\LambdaF}^{k} \LambdaF^{l}\bar{\LambdaF}_{k} \Lambda_{l} = 0\,. \label{eq:eomD}
	\end{align}
Note that both \eqref{eq:eomchi} and \eqref{eq:eomD} only depend on the coset scalars via $P_a$ and $\widetilde\FFF$ and are therefore manifestly $SU(1,1)$ invariant.

\subsection{Gauge fixing the R-symmetry group}

Before moving on to the derivation of the dilaton Weyl multiplet, we need to address an important point related to the $SU(4)$ R-symmetry group of $N=4$ conformal supergravity. As announced in the introduction, we will show in section~\ref{sec:construction} that the $N=4$ dilaton Weyl multiplet directly arises from the dimensional reduction of the six-dimensional $N=(2,0)$ Weyl multiplet. Since the six-dimensional R-symmetry is $USp(4)$, we anticipate here that it will be necessary to break the four-dimensional $SU(4)$ R-symmetry group to its $USp(4)$ subgroup in order to construct the dilaton Weyl multiplet. Note that this is a feature which is absent in the constructions of half-maximally supersymmetric dilaton Weyl multiplets \cite{BERGSHOEFF1986653,Bergshoeff:2001hc,Butter:2017pbp,Gold:2022bdk,Hutomo:2022hdi}. In these cases, the same R-symmetry group is indeed realized on both the standard Weyl and the dilaton Weyl multiplet. 

Let us then now discuss the gauge-fixing of the $SU(4)$ R-symmetry to $USp(4)$. We recall that $USp(4)$ is given as an intersection of Lie groups as $USp(4)\equiv Sp(4,\mathbb{C})\cap SU(4)$.\footnote{At the level of Lie algebras we also note the isomorphism $\mathfrak{usp}(4)\cong \mathfrak{so}(5)$, implying for instance that the ${\bf 5}$ representation of $USp(4)$ is real.} 
Hence apart from the invariant tensors $\varepsilon_{ijkl}$ and $\delta_{i}^{j}$ of $SU(4)$, it also has another invariant tensor $\Omega_{ij}$ coming from $Sp(4,\mathbb{C})$ which is an anti-symmetric $4\times 4$ matrix that we take as,
	\begin{align}
	    \Omega_{ij}=\begin{pmatrix}
	        0_{2\times 2} & \mathbb{1}_{2\times 2}\\-\mathbb{1}_{2\times 2}& 0_{2\times 2}
	    \end{pmatrix}\,.
	\end{align}
Its inverse $\Omega^{ij}$ is defined via the following relation
	\begin{align}
	    \Omega_{ij}\Omega^{kj}=\delta_{i}^{k}\,.
	\end{align}
	The $SU(4)$ invariant $\varepsilon_{ijkl}$ and the $Sp(4,\mathbb{C})$ invariant $\Omega_{ij}$ are related inside $USp(4)$ by
	\begin{align}\label{Omega_epsilon_relation}
	    \varepsilon_{ijkl}=-\Omega_{ij}\Omega_{kl}+\Omega_{ik}\Omega_{jl}-\Omega_{jk}\Omega_{il}\,.
	\end{align}
	The vector multiplet field $\phi^{ij}$, which transforms in the \textbf{6} of $SU(4)$, decomposes into a field $\mathring{\phiF}^{ij}$ in the \textbf{5} of $USp(4)$ and a $USp(4)$ singlet $\rho$, 
	\begin{align}
	    \phiF^{ij}=\mathring{\phiF}^{ij}+\frac{1}{4}\Omega^{ij}\rhoF\,.
     \label{eq:gaugefix}
	\end{align}
	Hence $\mathring{\phiF}^{ij}$ is $\Omega$-traceless, \textit{i.e} $\Omega_{ij}\mathring{\phiF}^{ij}=0$, and $\rhoF=\Omega_{ij}\phiF^{ij}$. The reality condition on $\phiF_{ij}$ (see Table~\ref{table:vmfields}), and the relation between the $USp(4)$ invariants given in~\eqref{Omega_epsilon_relation} imply that $\mathring{\phiF}_{ij}$ satisfies the same reality condition and that $\rhoF$ is real. Let us now gauge fix $SU(4)$ to $USp(4)$ by setting 
 \begin{align}
    \mathring\phiF_{ij} = 0\,.
    \label{eq:gfcondition}
    \end{align}
    This gauge fixing condition is not preserved by $Q$-supersymmetry transformations, which therefore have to be appropriately redefined by a compensating $SU(4)$ transformation,
	\begin{equation}
		\delta^{\text{new}}_Q(\epsilonF) \coloneqq  \delta_Q(\epsilonF) + \delta_{SU(4)}\big(k(\epsilon)\indices{^i_j}\big)\,.
		\label{eq:newsusy}
	\end{equation}
The field dependent parameter $k(\epsilon)\indices{^i_j}$ is determined by requiring $\delta^{\text{new}}_Q\mathring \phiF_{ij} = 0$. It reads
	\begin{align}\label{Udef}
	k(\epsilon)^{i}{}_{j} &=\frac{4}{\rhoF}\Omega^{ik}\left(\bar\epsilonF_{[j}\psiF_{k]} - \frac{1}{2}\varepsilon_{jklm}\bar\epsilonF^l\psiF^m - \text{$\Omega$-trace}\right)\nn\\*
&= - \frac{4}{\rho} \Omega_{jk}\left(\bar\epsilonF^{[i}\psiF^{k]} - \frac{1}{2}\varepsilon^{iklm}\bar\epsilonF_l\psiF_m -\Omega\text{-trace}\right)\,,
	\end{align}
and sits in the $\bf{5}$ of $USp(4)$ because of the subtraction of the $\Omega$-trace. For any object $O_{ij}$ carrying a pair of antisymmetric indices, the substraction of the $\Omega$-trace is given by
\begin{align}\label{Omegatrace}
    O_{ij}-\Omega\text{-trace}&=O_{ij}-\frac{1}{4}\Omega_{ij}\Omega^{kl}O_{kl}\,.
\end{align}
The subtraction of the $\delta$-trace, which will appear later on, is instead defined for any object $O^i{}_j$ carrying a pair of up and down indices as 
\begin{align}\label{deltatrace}
    O^{i}{}_{j}-\delta\text{-trace}&=O^{i}{}_{j}-\frac{1}{4}\delta^{i}_{j}O^{k}{}_{k}\,.
\end{align}
In the following, we will consistently work with the redefined transformations \eqref{eq:newsusy} and drop the superscript `new'. 

Now let us decompose the $SU(4)$ adjoint gauge field $V_a{}^i{}_j$ into fields transforming in the $\bf{10}\oplus\bf{5}$ of $USp(4)$ as
\begin{align}
    \VF_{a}{}^{i}{}_{k}\,\Omega^{jk}\eqqcolon \VF_{a}^{(ij)}+\VF_{a}^{[ij]}\,.
\end{align}
$\VF_{a}^{(ij)}$ will play the role of the $USp(4)$ gauge field. On the other hand, $\VF_{a}^{[ij]}$ which is in the $\bf{5}$ of $USp(4)$ should become a supercovariant field. However, due to the compensating $SU(4)$ transformation in \eqref{eq:newsusy}, the new $Q$-supersymmetry transformation of $\VF_{a}^{[ij]}$ involves terms proportional to $\partial_\mu\epsilonF$. These non-covariant terms can be absorbed into the following redefinition of $\VF_{a}^{[ij]}$,
\begin{align}
\label{Xdef}
    \XF_{a}{}^{ij}\coloneqq  \VF_{a}^{[ij]}-\frac{1}{2}k(\psiF_{a})^{[i}{}_{k}\Omega^{j]k}\;,
\end{align}
where the notation $k(\psiF_a)$ denotes the compensating parameter defined in (\ref{Udef}) in which the supersymmetry parameter $\epsilonF^i$ is replaced by the gravitino $\psiF_a^i$.  The redefined field $\XF_{a}{}^{ij}$ will be the one appearing in the dilaton Weyl multiplet as a supercovariant auxiliary field. In the following, we will then simply use $\VF_{a}^{ij}$ to denote the $USp(4)$ gauge field $\VF_{a}^{(ij)}$ without causing confusion. Note finally that the supercovariant derivative of a field decomposes into a new supercovariant derivative $D_a^\text{new}$, which is covariant with respect to the new $Q$-supersymmetry and $USp(4)$, plus some supercovariant terms that depend on $\XF_{a}^{ij}$. For instance for the gaugino $\psi_i$ appearing in Table~\ref{table:vmfields}, and its complex conjugate $\psi^i$, we have
\begin{subequations}
\begin{align}\label{USP4covder}
    D_{a}\psi_i&=D^\text{new}_{a}\psi_i+\XF_{a}{}^{jk}\psi_j\,\Omega_{ki}\,, \\
    D_{a}\psi^{i}&= D^\text{new}_{a}\psi^{i}-\XF_{a}{}^{ij}\psi^k\,\Omega_{jk}\,.
\end{align}
\end{subequations}
In the following sections, only $D^\text{new}_{a}$ will appear and we will again consistently drop the `new' superscript for convenience.

\subsection{\texorpdfstring{$N=4$ dilaton Weyl multiplet}{N=4 dilaton Weyl multiplet}}
\label{sec:dilWeylmu}

Once the $SU(4)$ gauge fixing condition \eqref{eq:gfcondition} has been imposed, the equations of motion for the vector multiplet that were presented in Section~\ref{sec:vmultiplet}
decompose according to certain $USp(4)$ representations.
This allows us to interpret these equations as constraints for some of the components of the auxiliary Weyl multiplet fields which then become composite. The remaining independent fields, including those of the vector multiplet, are ultimately gathered into a single new off-shell multiplet that is called the dilaton Weyl multiplet. We describe this procedure in detail below.

Consider first the auxiliary bosonic field $D^{ij}{}_{kl}$ of the Weyl multiplet which transforms in the $\bf{20'}$ of $SU(4)$. It decomposes into 
	\begin{align}
	    \DF^{ij}{}_{kl}=&\mathring{\DF}^{ij}{}_{kl}+\Omega^{ij}\mathring{\DF}_{kl}
     - \Omega_{kl} \Omega^{im} \Omega^{jn} \mathring{\DF}_{mn}
+\Omega^{ij}\Omega_{kl}\mathring{\DF}
     -\frac{2}{3}\delta^{[i}_{k}\delta^{j]}_{l}\mathring{\DF}\,,
	\end{align}
	where the fields $\mathring{\DF}^{ij}{}_{kl}$, $\mathring{\DF}_{ij}$ and $\mathring{\DF}$ transform in the $\bf{14}$, $\bf{5}$ and $\bf{1}$ of $USp(4)$, respectively, and satisfy the following identities under complex conjugation
 \begin{subequations}
	\begin{align}
 \label{D0def}
	    \left(\mathring{\DF}^{ij}{}_{kl}\right)^{*}
     &= \mathring{\DF}^{kl}{}_{ij}=\frac{1}{4}\varepsilon_{ijmn}\varepsilon^{klpq}\mathring{\DF}^{mn}{}_{pq}=\Omega^{kp}\Omega^{lq}\,\Omega_{im}\Omega_{jn}\mathring{\DF}^{mn}{}_{pq}\,, \\
	    \left(\mathring{\DF}_{ij}\right)^{*}&
     =-\frac{1}{2}\varepsilon^{ijkl}\mathring{\DF}_{kl}=-\Omega^{ik}\Omega^{jl}\mathring{\DF}_{kl}\,, \\
	    \mathring{\DF}^{*}&=\mathring{\DF}\,.
	\end{align}
 \end{subequations}
	Let us now rewrite the scalar field equation of motion \eqref{eq:eomD} by taking into account the gauge-fixing condition \eqref{eq:gfcondition}. This leads to
	\begin{align}
		&\frac{1}{4} \Omega_{ij}\, \Box\rhoF -D_{a}\rhoF\, \XF^{akl}\,\Omega_{ik}\Omega_{jl}-\frac{1}{2}\rhoF\, D_{a}\XF^{akl}\,\Omega_{ik}\Omega_{jl}+\rhoF \,\XF_{a}^{kl}\XF^{amn}\,\Omega_{km}\Omega_{il}\Omega_{jn}\nonumber \\
		& - (\TF_{ij}\cdot \widetilde{\FFF} - \frac{1}{2}\varepsilon_{ijkl}\,\TF^{kl}\cdot \widetilde{\FFF}) + (\bar\psiF_k\chiF\indices{^k_{ij}} - \frac{1}{2}\varepsilon_{ijkl}\,\bar\psiF^m\chiF\indices{_m^{kl}}) \nonumber\\
		&+ \frac{1}{2}\rhoF\left(\mathring{\DF}_{ij}+\frac{5}{6}\Omega_{ij}\mathring{\DF}\right) - \frac{1}{6}(2\bar \psiF_{[i} \slashed{\PF} \LambdaF_{j]} - \varepsilon_{ijkl}\, \bar \psiF^{k} \slashed{\bar{\PF}} \LambdaF^l) - \frac{1}{12}(2\bar \LambdaF^k\psiF_{[i}\EF_{j]k} - \varepsilon_{ijkl}\,\bar\LambdaF_m\psiF^k \EF^{lm})\nonumber\\
		&- \frac{1}{48}\rhoF\, \EF_{kl}\EF^{kl}\,\Omega_{ij} + \frac{1}{12}\rhoF \PF_{a}\bar{\PF}^{a}\,\Omega_{ij} + \frac{1}{48}\rhoF \,\Omega_{ij}(\bar \LambdaF^k\slashed{D} \LambdaF_k + \bar{\LambdaF}_{k}\slashed{D}\LambdaF^{k}-2\bar{\LambdaF}_{k}\slashed{\XF}^{kl}\LambdaF^{m}\,\Omega_{lm})\nonumber \\
		& + \frac{1}{32}\rhoF\,\Omega_{ij} \bar{\LambdaF}^{k} \LambdaF^{l}\bar{\LambdaF}_{k} \LambdaF_{l} = 0\,. \label{eq:eomDusp4}
	\end{align}
 We recall that all derivatives now include a covariantization with respect to the new $Q$-supersymmetry transformations \eqref{eq:newsusy} and $USp(4)$. We can now use projections of the above equation onto the $\bf{5}$ and $\bf{1}$ of $USp(4)$ to respectively eliminate $\mathring{\DF}_{ij}$ and $\mathring{\DF}$. Note that the field $\mathring \DF\indices{^{ij}_{kl}}$ simply dropped out in the field equation \eqref{eq:eomD} as a result of the gauge-fixing \eqref{eq:gfcondition}. It therefore remains as an independent field of the dilaton Weyl multiplet and we now simply relabel it as $\DF\indices{^{ij}_{kl}}$. 
 
Let us now consider the auxiliary fermion $\chiF_k{}^{ij}$ of the Weyl multiplet which transforms in the $\textbf{20}$ of $SU(4)$. It decomposes into the $\Omega$-trace ($\mathring\chiF_k$) and $\Omega$-trace-free ($\mathring\chiF_k{}^{ij}$) parts which respectively transform in the ${\textbf{4}}$ and $\textbf{16}$ of $USp(4)$,
	\begin{equation}
		\chiF_k{}^{ij} = \: \mathring\chiF_k{}^{ij} + \frac{1}{4}\Omega^{ij}\mathring\chiF_k+\frac{1}{6}\delta^{[i}_{k}\Omega^{j]l}\mathring{\chiF}_{l}\, .
	\end{equation}
	In the gauge \eqref{eq:gfcondition}, the equation of motion \eqref{eq:eomchi} for the vector multiplet gaugino $\psi_i$ takes the form 
	\begin{align}\label{eomchi_usp4}
		&\slashed{D}\psiF_i +\slashed{\XF}^{jk}\psi_{j}\Omega_{ki}+ \frac{1}{4}\gamma\cdot\widetilde{\FFF}\LambdaF_i + \frac{1}{2}\EF_{ij}\psiF^j -\frac{1}{4}\varepsilon_{ijkl}\gamma\cdot \TF^{jk}\psiF^l\nonumber \\
		&\hspace{10mm} -\frac{5}{24}\rhoF\mathring\chiF_i + \frac{1}{6}\Omega_{ij}\rhoF \EF^{jk}\LambdaF_k +\frac{1}{12} \slashed{\bar{\PF}}\Omega_{ij}\rhoF\LambdaF^j - \frac{1}{8}\gamma_a \psiF_j \bar\LambdaF_i\gamma^a\LambdaF^j = 0\,.
	\end{align}
	It can be used to eliminate the $\Omega$-trace part $\mathring{\chiF}_{k}$ from the off-shell multiplet in favor of the gaugino. Once again, the $\Omega$-traceless part $\mathring\chiF_k{}^{ij}$ remains as an independent field of the dilaton Weyl multiplet and we again simply relabel it as $\chiF_k{}^{ij}$. 
	
The only equation of motion that is left to be considered is the vector field equation~(\ref{eq:eomvec}), which can be written as
\begin{align}
D_{[a}\,G_{bc]}=0\,.
\end{align}
Following the logic of previous dilaton Weyl multiplet constructions \cite{Butter:2017pbp,BERGSHOEFF1986653,Bergshoeff:2001hc}, we now treat it as a Bianchi identity for the supercovariant field strength $G_{ab}$. In other words, we promote $G_{ab}$ to an independent field strength that derives from a new gauge field $B_\mu$. This means that we now have
\begin{equation}
G_{ab}=2\,e_{[a}{}^\mu e_{b]}{}^\nu \partial_{\mu} B_\nu+\text{gravitino terms}\,.\label{eq:GAnsatz}
\end{equation}
The next step is to interpret the equation \eqref{eq:G+} as a constraint that can be used to eliminate the $\Omega$-trace part of the auxiliary field $T_{abij}$ (and $T_{ab}{}^{ij}$) in favor $G^+_{ab}$ (and $G_{ab}^-$). To proceed, let us first decompose these auxiliary fields into the $\bf{5}$ ($\Omega$-traceless) and $\bf{1}$ ($\Omega$-trace) of $USp(4)$ as follows
\begin{subequations}
	\begin{align}
	    \TF_{abij}= \mathring{\TF}_{abij}+\frac{1}{4}\Omega_{ij}\mathring{\TF}
^{+}_{ab}\,, \\
	    \TF_{ab}{}^{ij}=\mathring{\TF}_{ab}{}^{ij}+\frac{1}{4}\Omega^{ij}\mathring{\TF}^{-}_{ab}\,.
	\end{align}
 \end{subequations}
The constraint (\ref{eq:G+}) and its complex conjugate then lead to
\begin{subequations}
\label{Tsol_both}
	\begin{align}
	    \mathring{\TF}^{+}_{ab}&=\frac{2}{\rhoF}\left[i\PhiF \,{\GF}^{+}_{ab}-\varphiF {\FFF}^{+}_{ab}+\frac{1}{2}\bar{\LambdaF}^{i}\gamma_{ab}\psiF_{i}\right]=\frac{4}{\rho}\left[\mathcal{F}_{ab;\alpha}^{+}\phi^{\alpha}+\frac{1}{4}\bar{\LambdaF}^{i}\gamma_{ab}\psiF_{i}\right]\;, \label{T_sol1} \\
	    \mathring{\TF}^{-}_{ab}&=\frac{2}{\rhoF}\left[-i\PhiF^{*} \,{\GF}^{-}_{ab}-\varphiF^{*} {\FFF}^{-}_{ab}+\frac{1}{2}\bar{\LambdaF}_{i}\gamma_{ab}\psiF^{i}\right]=\frac{4}{\rho}\left[\varepsilon^{\alpha\beta}\mathcal{F}_{ab;\alpha}^{-}\phi_{\beta}+\frac{1}{4}\bar{\LambdaF}_{i}\gamma_{ab}\psiF^{i}\right]\,,\label{T_sol2}
	\end{align}
 \end{subequations}
where the $SU(1,1)$ doublet $\mathcal{F}_{ab;\alpha}$ was defined in \eqref{F12def}. This allows us to trade-off the off-shell field $\mathring{\TF}^{\pm}_{ab}$ for the field strength $G_{ab}^{\pm}$. As a result, the two vector fields $A_\mu$ and $B_\mu$ are promoted to off-shell fields, and they both become part of the dilaton Weyl multiplet. In the following, we will once again relabel the remaining components $\mathring{\TF}_{ab}{}^{ij}$ and $\mathring{\TF}_{abij}$ as $\TF_{ab}{}^{ij}$ and $\TF_{abij}$.

 Let us now introduce the $SU(1,1)$ covariant gauge field $\mathcal A_{\mu;\alpha}$ that is associated with the field strength $\mathcal F_{ab;\alpha}$. Its components are given by
\begin{align}
    \mathcal{\AF}_{\mu; 1} \coloneqq -\frac{1}{2}\left( \AF_\mu - i B_\mu \right)\,,\quad \mathcal{\AF}_{\mu; 2} \coloneqq -\left(\mathcal{\AF}_{\mu; 1}\right)^*\,,
\end{align} 
and therefore satisfy $\left(\mathcal{A}_{\mu;\alpha}\right)^{*}=\varepsilon_{\alpha\beta}\eta^{\beta\gamma}\mathcal{A}_{\mu;\gamma}$. Under $g\in SU(1,1)$, they must transform as
\begin{align}
    \begin{pmatrix} {\mathcal{A}}_{\mu;1} \\{\mathcal{A}}_{\mu;2}\end{pmatrix}\to g \begin{pmatrix} {\mathcal{A}}_{\mu;1}\\{\mathcal{A}}_{\mu;2}\end{pmatrix}\,,\label{GaugeFieldUtransf}
\end{align}
in order to be consistent with the transformation \eqref{FieldSUtransf} of $\mathcal{F}_{ab;\alpha}$. Since $SU(1,1)$ commutes with supersymmetry, the property \eqref{GaugeFieldUtransf} and the knowledge of the supersymmetry variation of $A_\mu$(which can be found in \cite{deRoo:1984zyh}), allows us to directly compute the supersymmetry variation of $B_\mu$. The result is presented below in \eqref{eq:SUSYcA}. Note that the supersymmetry variation of $B_\mu$ can also be extracted from the variation of $G_{ab}$, which can itself be computed from the constraints \eqref{Tsol_both}. The supercovariance of the Ansatz \eqref{eq:GAnsatz} for $G_{ab}$ indeed implies that in its variation, all the terms that contain a spacetime derivative acting on the fields (including the curvatures) originate from the variation of $B_\mu$.

 We have now treated all the equations of motion of the vector multiplet as constraints for some of the components of the auxiliary Weyl mulitplet fields. The remaining set of independent fields is summarized in Table~\ref{table:dilaton_Weyl}, and defines the off-shell $N=4$ dilaton Weyl multiplet. 
\begin{table}[t!]
 \begin{center}
   \scalebox{0.80}{
 \begin{tabular}{ c | c c l c c c }
 \hline 
 \hline
   & Field & Gauge symmetry & Name/Restrictions & $USp(4)$ & $w$ & $\rm c$ \\
 \hline
 \multirow{14}{*}{Bosons} & $\eF_\mu{}^a$ & Translations & vierbein & $\mathbf{1}$ & $-1$ & 0 \\
 & $\omegaF_{\mu}{}^{ab}$  & local Lorentz & spin connection & $\mathbf{1}$ & 0 & 0 \\
 & $\bbF_{\mu}$ & Dilatations & dilatational gauge field & $\mathbf{1}$ & 0 & 0 \\
 & $\VF_{\mu}{}^{ij}$ & USp(4) & $\mathrm{USp}(4)$ gauge field; $\VF_\mu{}^{ij}=\VF_{\mu}{}^{ji}$  & $\mathbf{10}$ & 0 & 0 \\
 & & &  $\VF_{\mu ij}\equiv (\VF_{\mu}^{ij}){}^{*}=\VF_{\mu}{}^{kl}\Omega_{ik}\Omega_{jl}$  & & & \\
 & $\ffF_{\mu}{}^a$ & conformal boosts & K-gauge field & $\mathbf{1}$ & 1 & 0 \\
 & $\aaF_\mu$ & U(1)& $\mathrm{U}(1)$ gauge field & $\mathbf{1}$ & 0 & 0 \\
 & $\mathcal{A}_{\mu;\alpha}$ & U(1) & U(1) gauge fields, SU(1,1) doublet $\alpha=1,2$  & $\mathbf{1}$ & 0 & 0 \\
 & $\XF_{a}{}^{ij}$ & & $\XF_{a}{}^{ij}=-\XF_{a}{}^{ji}$, $\Omega_{ij}\XF_{a}{}^{ij}=0$ & $\mathbf{5}$ & 1 & 0 \\
 & & & $\XF_{a ij}\equiv (\XF_{a}^{ij}){}^{*}=-\XF_{a}{}^{kl}\Omega_{ik}\Omega_{jl}$  & &  & \\
 & $\rhoF$ & & $\rhoF=\rhoF^{*}$ & $\mathbf{1}$ & 1 & 0 \\
 & $\phiF_{\alpha}$ & & $\phiF_\alpha\,\phiF^{\alpha}=1\,,\,\phiF^1=\phiF_1^*\,,\phiF^2=-\phiF_2^*$ & $\mathbf{1}$ & 0 & $-1$ \\
 & $\EF_{ij}$ & & $\EF_{ij}=\EF_{ji}$ & ${\mathbf{10}}$ & 1 & $-1$ \\
 &$\TF_{ab}{}^{ij}$ & & $\tfrac12\varepsilon_{ab}{}^{cd}\TF_{cd}{}^{ij}=-\TF_{ab}{}^{ij}$ & $\mathbf{5}$ & 1 & $-1$\\
 & & & $\TF_{ab}{}^{ij}=-\TF_{ab}{}^{ji}$, $\Omega_{ij}\TF_{ab}{}^{ij}=0$ & & & \\
 & $\DF^{ij}{}_{kl}$ & & $\DF^{ij}{}_{kl}=\tfrac14 \varepsilon^{ijmn}\varepsilon_{klpq}\DF^{pq}{}_{mn}=\Omega^{im}\Omega^{jn}\Omega_{kp}\Omega_{lq}\DF^{pq}{}_{mn}$ & $\mathbf{14}$ & 2 & $0$ \\
 & & & $\DF_{kl}{}^{ij}\equiv(\DF^{kl}{}_{ij})^\ast=\DF^{ij}{}_{kl}$ & & & \\
 & & & $\DF^{ij}{}_{kj}=0$, $\Omega_{ij}\DF^{ij}{}_{kl}=\Omega^{kl}\DF^{ij}{}_{kl}=0$ & & & \\[1mm]
 \hline
\multirow{4}{*}{Fermions} 
 & $\phiF_{\mu\,i}$ & $S$-supersymmetry & $S$-gauge field; $\ga_*\,\phiF_{\mu i}=\phiF_{\mu i}$ & ${\mathbf{4}}$ & $\tfrac12$ & $\frac{1}{2}$ \\
& $\psiF_{\mu}{}^i$ & $Q$-supersymmetry & gravitini; $\ga_*\,\psiF_{\mu}{}^i=\psiF_{\mu}{}^i$ & $\mathbf{4}$ & $-\frac12$ & $-\frac12$ \\
& $\LambdaF_i$  & & $\ga_*\LambdaF_i=\LambdaF_i$ & ${\mathbf{4}}$ & $\tfrac{1}{2}$ & $-\tfrac32$ \\
& $\chiF_k{}^{ij}$  & & $\ga_*\chiF_k{}^{ij}=\chiF_k{}^{ij}$; $\chiF_k{}^{ij} =-\chiF_k{}^{ji}$ & $\mathbf{16}$ & $\tfrac32$ & $-\tfrac12$ \\[1mm]
& & & $\chiF_j{}^{ij}=0$, $\Omega_{ij}\chiF_k{}^{ij} =0$ & & & \\
& $\psiF_i$  & & $\ga_*\psiF_i=-\psiF_i$ & ${\mathbf{4}}$ & $\tfrac{3}{2}$ & $-\tfrac12$ \\
 \hline
 \hline
\end{tabular}}
$ $\newline
\caption{\textit{Fields of the $N=4$ dilaton Weyl multiplet.}}
\label{table:dilaton_Weyl}
\end{center}
\end{table}
Its supersymmetry transformations can be directly computed from the supersymmetry transformations of the standard Weyl and vector multiplets, by taking into account the $SU(4)$ compensating transformation  (\ref{eq:newsusy}) due to the R-symmetry gauge-fixing. The supersymmetry transformations of the dilaton Weyl multiplet fields then read,
\begin{subequations}
   \label{dil_weyl_trans}
\begin{align}
		\delta \eF^a_\mu = & \: \bar{\epsilonF}^i \gamma^a \psiF_{\mu i}  + \text{h.c.}\,,
  \\[5pt]
		\delta \psiF_\mu^i = &\: 2\mathcal{D}_\mu \epsilonF^i - 2\XF\indices{_\mu^i^k}\Omega_{kj}\epsilonF^j - \frac{1}{2}\gamma^{ab}\left(\TF\indices{_{ab}^{ij}} + \frac{1}{4}\Omega^{ij}\mathring{\TF}_{ab}^-\right)\gamma_\mu \epsilonF_j + \varepsilon^{ijkl}\bar{\psiF}_{\mu j} \epsilonF_k\LambdaF_l \nonumber \\
  &+ k(\epsilonF)\indices{^i_j}\psiF_\mu^j  - k(\psiF_\mu)\indices{^i_j}\epsilonF^j\,, \label{eq:dwgrav}
  \\[5pt]
  \delta \bbF_\mu = &\: \frac{1}{2}\bar{\epsilonF}^i\phiF_{\mu i} + \text{h.c.}\,, \label{eq:SUSYb}
  \\[5pt]
		\delta \VF_\mu^{ij} = & \:\bar{\epsilonF}^{(i}\phiF_{\mu k} \Omega^{j)k} + \bar{\epsilonF}^s\gamma_\mu \chiF\indices{^{(i}_{sk}}\Omega^{j)k} -\frac{1}{2}\varepsilon_{klsn}\EF^{l(i}\Omega^{j)k}\bar{\epsilonF}^s\psiF_\mu^n -\frac{1}{6}\EF^{l(i}\bar{\epsilonF}_k\gamma_\mu \LambdaF_l \Omega^{j)k} \nonumber 
  \\
		&+ \frac{1}{3}\bar\epsilonF^{(i}\gamma_\mu \slashed{\PF}\LambdaF_k\Omega^{j)k} - \frac{1}{4} \Omega^{k(i}\varepsilon^{j)sln}\TF\indices{^{ab}_{lk}}\bar{\epsilonF}_s\gamma_{ab}\gamma_\mu \LambdaF_n\nonumber \\
  &+ \frac{1}{4}\Omega^{k(i}\varepsilon^{j)slp}\varepsilon_{ktnp}\bar{\epsilonF}^t\gamma_a \psiF_{\mu s}\bar{\LambdaF}_l\gamma^a\LambdaF^n+\text{h.c.}  
  \,,\label{eq:SUSYV}\\[5pt]	
		\delta \mathcal{\AF}_{\mu;\alpha} = & \: \varepsilon_{\beta\alpha}\phiF^\beta \left( \bar{\epsilonF}^i\gamma_\mu\psiF_i -{\frac{1}{2}}\rhoF\bar{\epsilonF}^i\psiF_\mu^j \Omega_{ij} {+} \frac{\rhoF}{{4}} \bar{\epsilonF}_i\gamma_\mu\LambdaF_j\Omega^{ij} \right) \nonumber\\[4pt]
  &\,- \phiF_\alpha\left( \bar{\epsilonF}_i\gamma_\mu\psiF^i -{\frac{1}{2}}\rhoF\bar{\epsilonF}_i\psiF_{\mu\,j} \Omega^{ij} {+} \frac{\rhoF}{{4}} \bar{\epsilonF}^i\gamma_\mu\LambdaF^j\Omega_{ij} \right)\,, \label{eq:SUSYcA}
  \\[5pt]
		\delta \rhoF = &\: {2}\Omega^{ij}\bar{\epsilonF}_i\psiF_j + \text{h.c.}\,,\label{eq:SUSYrho}
  \\[5pt]
		\delta \phiF^\alpha = & \:\bar{\epsilonF}_i\LambdaF^i \varepsilon^{\alpha\beta}\phiF_\beta\,,\label{eq:susyphia}
  \\
		\delta \psiF_i = &\: -\frac{1}{2}\gamma_{ab}\epsilonF_i \widetilde{\FFF}^{ab+} - {\frac{1}{2}}\Omega_{ij}(\slashed{D}\rhoF)\epsilonF^j + \rhoF \XF_{a\,ij}\gamma^a\epsilonF^j +{\frac{\rhoF}{4}} \EF_{ij}\Omega^{jk}\epsilonF_k + \frac{1}{2}\bar\LambdaF_j\psiF^j\epsilonF_i -\bar\LambdaF_i\psiF^j\epsilonF_j \nonumber\\*
		&+ \frac{1}{{8}}\rhoF\Omega_{kl}\bar\LambdaF^l\gamma_a\LambdaF_i\gamma^a\epsilonF^k - k^k{}_i(\epsilon)\psiF_k\,,\label{eq:SUSYpsi} 
  \\[5pt]
		 \delta \LambdaF_i = &-2\bar{\slashed{\PF}}\epsilonF_i + \EF_{ij}\epsilonF^j + \frac{1}{2}\varepsilon_{ijkl}\left(\TF\indices{_{ab}^{kl}} + \frac{1}{4}\Omega^{kl}\mathring{\TF}^-_{ab}\right)\gamma^{ab}\epsilonF^j -k(\epsilon)^k{}_i\,\LambdaF_k\,,
   \label{eq:dwLambda}
   \\[5pt]
  \delta \XF\indices{_a ^{ij}} = &\: \bar\epsilonF^s\gamma_a \chiF\indices{^{[i}_{sk}}\Omega^{j]k} -\frac{1}{3}\bar{\epsilon}^{[i}\gamma_{a}\mathring{\chi}^{j]}-\frac{1}{6}\EF^{l[i}\bar\epsilonF_k\gamma_a\LambdaF_l\Omega^{j]k} + \frac{1}{3}\bar\epsilonF^{[i}\gamma_a\slashed{\PF}\LambdaF_k\Omega^{j]k}\nonumber \\
      &- \frac{1}{4}\varepsilon^{sln[i}\bar\epsilonF_s\left( \TF_{cd\,lk} + \frac{1}{4}\Omega_{lk}\mathring{\TF}^+_{cd}\right) \gamma^{cd} \, \gamma_a \LambdaF_n\Omega^{j]k}- \frac{{4}}{\rhoF}\Omega^{s[i}\bar\epsilonF_{{s}}\psiF_{{k}} \XF\indices{_a^{j]k}} 
      \\
		&  -{2}\varepsilon^{ijkl}\bar\epsilonF_k D_a \left(\frac{\psiF_l}{\rhoF}\right) + \frac{1}{8}\varepsilon^{ijkl}\bar\epsilonF^m\gamma_a \gamma^{bc}\psiF_l \left(\TF_{bc\,mk} + \frac{1}{4}\Omega_{mk}\mathring{\TF}_{bc}^+ \right)- \text{$\Omega$-trace} -\text{h.c.}\,,\nonumber\\[5pt]
		\delta \EF_{ij} = &\: 2\bar{\epsilonF}_{(i}\slashed{D}\LambdaF_{j)}+2\Omega_{l(j}\bar{\epsilonF}_{i)}\slashed{\XF}^{kl}\LambdaF_{k} - 2\bar{\epsilonF}^k\chiF_{(i}{}^{mn}\varepsilon_{j)kmn} +  \bar\epsilonF^k\mathring{\chiF}_{(i}\Omega_{j)k} - \bar{\LambdaF}_i\LambdaF_j\bar{\epsilonF}_k\LambdaF^k + 2\bar{\LambdaF}_k\LambdaF_{(i}\bar{\epsilonF}_{j)}\LambdaF^k \nonumber \\[2pt]
         &-2\,k(\epsilon)^k{}_{(i}\,\EF_{kj)}\,,
         \\[5pt]
		\delta \TF\indices{_{ab}^{ij}} = &\: 2\bar{\epsilonF}^{[i}R(Q)\indices{_{ab}^{j]}} + \frac{1}{2}\bar{\epsilon}^k\gamma_{ab}\chiF_k{}^{ij} + \frac{1}{4}\varepsilon^{ijkl}\bar{\epsilonF}_k\gamma^c\gamma_{ab} \DF_c\Lambda_l+\frac{1}{4}\varepsilon^{ijkl}\bar{\epsilonF}_k\gamma^c\gamma_{ab} \Lambda_m X_{c}{}^{mn}\Omega_{nl}\nonumber \\ 
		&  - \frac{1}{6}\EF^{k[i}\bar{\epsilonF}^{j]}\gamma_{ab}\LambdaF_k + \frac{1}{3}\bar{\epsilonF}^{[i}\gamma_{ab}\bar{\slashed{\PF}}\LambdaF^{j]}+ 2\,k(\epsilon)^{[i}{}_k\,\TF\indices{_{ab}^{kj]}} - \text{$\Omega$-trace}\,,
  \\[5pt]
		\delta \chiF_k{}^{ij} = &\:-\frac{1}{2}\gamma^{ab}\big(\slashed{D}\TF\indices{_{ab}^{ij}}\big) \epsilonF_k-\gamma^{ab}\Omega_{ml}\slashed{X}^{l[i}T_{ab}{}^{j]m}\epsilon_{k} - \gamma^{ab}\Omega_{kl}R(\VF)\indices{_{ab}^{l[i}}\epsilonF^{j]}-2\Omega_{kl}D_{a}X_{b}{}^{l[i}\gamma^{ab}\epsilon^{j]}\nonumber \\
     &+2X_{a}{}^{l[i}X_{b}{}_{lk}\gamma^{ab}\epsilon^{j]} - \frac{1}{2}\varepsilon^{ijlm}\big(\slashed{D}\EF_{kl}\big)\epsilonF_m-\varepsilon^{ijlm}\Omega_{p(l}E_{k)n}\slashed{X}^{np}\epsilon_{m} + \DF\indices{^{ij}_{kl}}\epsilonF^l \nonumber\\
		& -\frac{1}{6}\varepsilon_{klmn}\EF^{l[i}\gamma^{ab}(\TF\indices{_{ab}^{j]n}}\epsilonF^m + \TF\indices{_{ab}^{mn}}\epsilonF^{j]} + \frac{1}{4}\Omega^{j]n}\mathring{\TF}_{ab}^{-}\epsilonF^m + \frac{1}{4}\Omega^{mn}\mathring{\TF}_{ab}^{-}\epsilonF^{j]}) + \frac{1}{2}\EF_{kl}\EF^{l[i}\epsilonF^{j]} \nonumber\\
		& - \frac{1}{2}\varepsilon^{ijlm}\bar{\slashed{\PF}}\gamma_{ab}\left(\TF\indices{^{ab}_{kl}} + \frac{1}{4}\Omega_{kl}\mathring{\TF}^{ab+}\right)\epsilonF_m + \frac{1}{4}\gamma^a\epsilonF_n(2\varepsilon^{ijln}\bar{\chiF}\indices{^m_{lk}} - \varepsilon^{ijlm}\bar{\chiF}\indices{^n_{lk}})\gamma_a\LambdaF_m \nonumber \\
      &+ \frac{1}{4}\epsilonF^{[i}\left(2\bar{\LambdaF}^{j]}\slashed{D}\LambdaF_k + \bar{\LambdaF}_k\slashed{D}\LambdaF^{j]}\right) + \frac{1}{4}\epsilonF^{[i}\left(2\bar{\LambdaF}^{j]}\slashed{X}^{lm}\Omega_{mk}\LambdaF_l - \bar{\LambdaF}_k\slashed{X}^{j]m}\Omega_{ml}\LambdaF^{l}\right)\nonumber \\
      & -\frac{1}{4}\gamma^{ab}\epsilonF^{[i}(2\bar{\LambdaF}^{j]}\gamma_a D_b\LambdaF_k - \bar{\LambdaF}_k\gamma_a D_b\LambdaF^{j]})-\frac{1}{4}\gamma^{ab}\epsilonF^{[i}(2\bar{\LambdaF}^{j]}\gamma_a X_{b}{}^{lm}\Omega_{mk}\LambdaF_l + \bar{\LambdaF}_k\gamma_a X_{b}{}^{j]m}\Omega_{ml}\LambdaF^{l}) \nonumber\\
		& - \frac{5}{12}\varepsilon^{ijlm}\LambdaF_m \bar{\epsilonF}_l(\EF_{kn}\LambdaF^n - 2\slashed{\PF}\LambdaF_k)+ \frac{1}{12}\varepsilon^{ijlm}\LambdaF_m\bar{\epsilonF}_k(\EF_{ln}\LambdaF^n - 2\slashed{\PF}\LambdaF_l)  \nonumber \\
    & - \frac{1}{2}\gamma^{ab}\TF\indices{_{ab}^{l[i}}\gamma^c\epsilonF_{[k}\bar{\LambdaF}^{j]}\gamma_c\LambdaF_{l]} - \frac{1}{2}\gamma^{ab}\TF\indices{_{ab}^{ij}}\gamma^c\epsilonF_{[k}\bar{\LambdaF}^l\gamma_c\LambdaF_{l]} + \frac{1}{2}\epsilon^{[i}\bar{\LambdaF}^{j]}\LambdaF^m\bar{\LambdaF}_k\LambdaF_m  \nonumber\\[2pt]
   &- k(\epsilon)^t{}_k\,\chiF_t{}^{ij}
   - 2\,k(\epsilon)^{[i}{}_t\,\chiF_k{}^{j]t}- \Omega\text{-trace}-\text{$\delta$-trace}\,,
   \\[5pt]
		 \delta \DF\indices{^{ij}_{kl}} = &\:-4\bar{\epsilonF}^{[i}\slashed{D}\chiF\indices{^{j]}_{kl}}+4\bar{\epsilonF}^{[i}\slashed{X}^{j]n}\Omega_{nm}\chiF^{m}{}_{kl}-8\bar{\epsilonF}^{[i}\gamma^{a}\chiF^{j]}{}_{m[k}X_{a}{}^{mn}\Omega_{l]n}\nonumber \\[2pt]
   & + \varepsilon_{klmn}\bar{\epsilonF}^{[i}\left(- \frac{1}{2}\gamma^{ab}\gamma^{c}\LambdaF^{j]}\overleftrightarrow{D_{c}}\TF\indices{_{ab}^{mn}}-\frac{1}{2}\gamma^{ab}\slashed{X}^{j]q}\Omega_{qp}\LambdaF^{p}T_{ab}{}^{mn}-\gamma^{ab}\gamma^{c}\Lambda^{j]}T_{ab}{}^{np}{X}_{c}{}^{mq}\Omega_{qp} \right. \nonumber \\*
		& \;\;\;\;\;\;\;\;\;\;\;\; \;\;\;\;\;\; \left.-2 \EF^{j]p}\chiF_p{}^{mn} + \frac{1}{3}\EF^{j]m}\EF^{np}\LambdaF_p-\frac{2}{3}E^{j]n}\bar{\slashed{\PF}}\LambdaF^m  + \frac{1}{2}\gamma^{ab}\LambdaF_p \bar{\LambdaF}^{j]}\LambdaF^p\TF\indices{_{ab}^{mn}}\right)\nonumber \\*
  & + \varepsilon^{ijmn}\bar{\epsilonF}^p \TF\indices{^{ab}_{kl}}\left(2 \TF_{ab\,np}\LambdaF_m + \TF_{ab\,mn}\LambdaF_p\right)-2\bar{\LambdaF}^{[i}\gamma_a\LambdaF_m\bar{\epsilonF}^{j]}\gamma^a\chiF\indices{^m_{kl}} \nonumber\\*
		& + \bar{\epsilonF}^{[i}\left( 2 \bar{\slashed{\PF}}\gamma_{ab}\TF\indices{^{ab}_{kl}}\LambdaF^{j]} + \frac{2}{3}\LambdaF_{[k}\EF_{l]m}\bar{\LambdaF}^{j]}\LambdaF^m + \frac{1}{6}\gamma^{ab}\slashed{\PF}\LambdaF^{j]}\bar{\LambdaF}_k\gamma_{ab}\LambdaF_l\right) \nonumber\\*
		& - 2\,k(\epsilon)^{[i}{}_t\,\DF\indices{^{j]t}_{kl}} + 2\,k(\epsilon)^t{}_{[k}\,\DF\indices{^{ij}_{l]t}} + \text{h.c.}- \Omega\text{-trace}-\delta\text{-trace}\,,
	\end{align}
\end{subequations} 
Note that in the above transformation rules, $\mathring{T}^{\pm}_{ab}$ and $\mathring{\chiF}_i$ are still appearing in several places. These are now composite fields whose expressions in terms of the fundamental fields of the dilaton Weyl multiplet can be straightforwardly extracted from~\eqref{Tsol_both} and~\eqref{eomchi_usp4}. We chose not to substitute them here for the sake of brevity. Note also that the other composite fields $\mathring{D}_{ij}$ and $\mathring{D}$, whose expressions follow from \eqref{eq:eomDusp4}, simply drop out from the supersymmetry transformations. The subtraction of the $\Omega$- and $\delta$-traces that appear in some the above transformations were defined in (\ref{Omegatrace}) and (\ref{deltatrace}), while the expression of the compensating $SU(4)$ transformation parameter $k$ is given by \eqref{Udef}. Let us also define the Hermitian conjugation that appears above. For a generic term with $n$ upper and $m$ lower $USp(4)$ indices, denoted here schematically by $\phi^{i_1 i_2\cdots i_n}_{j_1 j_2\cdots j_m}$, we have,
\begin{align}
\phi^{i_1 i_2\cdots i_n}_{j_1 j_2\cdots j_m}\pm\text{h.c.}=\phi^{i_1 i_2\cdots i_n}_{j_1 j_2\cdots j_m}\pm\Omega^{i_1k_1}\cdots \Omega^{i_nk_n}\Omega_{j_1l_1}\cdots \Omega_{j_ml_m}
(\phi^{k_1 k_2\cdots k_n}_{l_1 l_2\cdots l_m})^{\dagger}\,,
\end{align}
where we recall that with our chiral notation $(\phi^{k_1 k_2\cdots k_n}_{l_1 l_2\cdots l_m})^{\dagger}=\phi^{l_1 l_2\cdots l_m}_{k_1 k_2\cdots k_n}$. 

The $SU(1,1)$ invariant $\tilde{F}_{ab}$ appearing in the variation \eqref{eq:SUSYpsi} was defined in~\eqref{FFF}. It combines the coset scalars $\phi^\alpha$ together with the supercovariant field strength $\mathcal{F}_{ab;\alpha}$. The complete expression of the latter can be determined from the supersymmetry transformation \eqref{eq:SUSYcA} of the associated gauge field $\mathcal{A}_{\mu;\alpha}$, and is given by
\begin{align}
\label{Falphadef}
{\mathcal{F}}_{ab;\alpha}=2\,e^{\mu}_{a}e^{\nu}_{b}\partial_{[\mu}\mathcal{A}_{\nu];\alpha}&+\varepsilon_{\alpha\beta}\phi^{\beta}\left(\bar{\psi}_{[a}^{i}\gamma_{b]}\psi_{i}-\frac{\rho}{{4}}\bar{\psi}_{a}^{i}\psi_{b}^{j}\Omega_{ij}+\frac{\rho}{{4}} \bar{\psi}_{i[a}\gamma_{b]}\Lambda_{j}\Omega^{ij}\right)\nonumber \\*
     & +\phi_{\alpha}\left(\bar{\psi}_{i[a}\gamma_{b]}\psi^{i}-\frac{\rho}{{4}}\bar{\psi}_{a i}\psi_{b j}\Omega^{ij}+\frac{\rho}{{4}} \bar{\psi}_{[a}^{i}\gamma_{b]}\Lambda^{j}\Omega_{ij}\right)\;.
 \end{align}
This is in line with the Ansatz \eqref{eq:GAnsatz} for the field strength $G_{ab}$, and with the $SU(1,1)$ transformations \eqref{FieldSUtransf}, \eqref{GaugeFieldUtransf} and \eqref{coset_transf}. The supersymmetry transformations (\ref{dil_weyl_trans}) also involve the supercovariant curvatures $R(Q)_{ab}^{i}$ and $R(V)_{ab}{}^{ij}$, which are associated to $Q$-supersymmetry and $USp(4)$ R-symmetry. Their expressions can again be deduced from the supersymmetry transformations of the corresponding gauge fields, and read
 \begin{align}\label{curvatures}
 R(V)_{ab}{}^{ij}=\;&2e^{\mu}_{a}e^{\nu}_{b}\partial_{[\mu}V_{\nu]}^{ij}-2V_{[a}^{ik}V_{b]}^{jl}\Omega_{kl}\nonumber \\
  +&\left[-\bar{\psiF}_{[a}^{(i}\phiF_{b] k}^{\phantom{i)}} \Omega^{j)k} - \bar{\psiF}_{[a}^s\gamma_{b]}^{\phantom{s}} \chiF\indices{^{(i}_{sk}}\Omega^{j)k} +\frac{1}{4}\varepsilon_{klsn}\EF^{l(i}\Omega^{j)k}\bar{\psiF}_{a}^s\psiF_b^n +\frac{1}{6}\EF^{(il}\bar{\psiF}_{k[a}\gamma_{b]} \LambdaF_l \Omega^{j)k}\right. \nonumber \\
		&- \frac{1}{3}\bar\psiF_{[a}^{(i}\gamma_{b]}^{\phantom{i)}} \slashed{\PF}\LambdaF_k\Omega^{j)k} - \frac{1}{4}\varepsilon^{(isln}\TF\indices{_{cdlk}}\bar{\psiF}_{s[a}\gamma^{cd}\gamma_{b]} \LambdaF_n\Omega^{j)k} \nonumber \\
  & \left.+ \frac{1}{8}\varepsilon^{(islp}\varepsilon_{ktnp}\bar{\psiF}_{[a}^t\gamma^{c} \psiF_{b] s}\bar{\LambdaF}_l\gamma_a\LambdaF^n\Omega^{j)k}+\text{h.c.}\right] \,, \\			
         R(Q)_{ab}^{i}=\; & 2e^{\mu}_{a}e^{\nu}_{b}\mathcal{D}_{[\mu}\psi_{\nu]}^{i}-\gamma_{[a}\phi^{i}_{b]}-2X_{[a}^{ik}\psi_{b]}^{j}\Omega_{kj}-\frac{1}{2}\gamma^{cd}\left(T_{cd}{}^{ij}+\frac{1}{4}\Omega^{ij}\mathring{T}_{cd}^{-}\right)\gamma_{[a}\psi_{b]j}\nonumber \\
         & +\frac{1}{2}\varepsilon^{ijkl}\bar{\psi}_{aj}\psi_{bk}\Lambda_{l}-k(\psi_{[a})^{i}{}_{j}\psi_{b]}^{j}\,.
 \end{align} 
Note that the dependent $S$-supersymmetry gauge field $\phi_{\mu\,i}$, which in particular appears in the transformations \eqref{eq:SUSYb} and \eqref{eq:SUSYV}, is determined via the curvature constraint $\gamma^a R(Q)^i_{ab}=0$.
  
 To conclude this section, we present the (non-vanishing) $S$-supersymmetry transformation, with parameter $\etaF^i$, of the dilaton Weyl multiplet fields,
 \begin{subequations}
	\begin{align}
		\delta \psiF_\mu^i = &\: - \gamma_\mu \etaF^i\,,\\
		\delta \bbF_\mu = &\: -\frac{1}{2}\bar{\psiF}_\mu^i\etaF_i + \text{h.c.}\,,\\
		\delta \VF_\mu^{ij} = &\: - \bar{\psiF}_\mu^{(i}\etaF_k\Omega^{j)k} + \text{h.c.}\,,\\
		\delta \psiF_i = &\: -\frac{\rhoF}{{2}}\,\Omega_{ij}\etaF^j\,,\\
		\delta \XF\indices{_a ^{ij}} = &\: -\frac{{2}}{\rhoF} \left( \bar{\etaF}^{[i}\gamma_a \psiF^{j]} - \frac{1}{2}\varepsilon^{ijkl}\bar{\etaF}_k\gamma_a\psiF_l - \text{$\Omega$-trace}\right)\,,\\
		\delta \EF_{ij} = &\: 2\bar\etaF_{(i}\LambdaF_{j)}\,,\\
		\delta \TF\indices{_{ab}^{ij}} = &\: - \frac{1}{4}\varepsilon^{ijkl}\bar\etaF_k\gamma_{ab}\LambdaF_l - \text{$\Omega$-trace}\,,\\
		\delta \chiF_k{}^{ij} = &\: \frac{1}{2}\TF\indices{_{ab}^{ij}}\gamma^{ab}\etaF_k  - \frac{1}{2}\varepsilon^{ijlm}\EF_{kl}\etaF_m - \frac{1}{4}\bar{\LambdaF}_k\gamma^a\LambdaF^{[i}\gamma_a\etaF^{j]} -  \text{$\Omega$-trace}-\text{$\delta$-trace}\,.
	\end{align}
 \end{subequations}
Note that the $SU(4)$ gauge-fixing condition~\eqref{eq:gfcondition} is inert under $S$-supersymmetry, so that there is no compensating transformation needed. The equations above follow directly from the Weyl and vector multiplet expressions with the exception of $\delta \XF_a{}^{ij}$ for which one has to use the definition~\eqref{Xdef}.

Let us finally point out that the existence of the $N=4$ dilaton Weyl multiplet, which only realizes an $USp(4)$ R-symmetry, might appear in contradiction with Nahm's classification of rigid superconformal algebras~\cite{Nahm:1977tg}. The latter indeed always involve an $SU(4)$ R-symmetry group. Upon closer inspection however, one can show that the rigid limits of the soft superconformal algebras\footnote{We recall that soft algebras are the ones appearing in (conformal) supergravities theories and for which the structure constants are field-dependent.} which are realized on dilaton Weyl multiplets lead to Poincar{\'e} (rather than superconformal) superalgebras, and therefore do not conflict with Nahm's classification. This was discussed for the case of five-dimensional $N=2$ conformal supergravity in \cite{Adhikari:2023tzi}.

\section{Dimensional reduction from six dimensions}
	\label{sec:dw6d}
	
In this section, we relate the fields of the four-dimensional dilaton Weyl multiplet to those of the six-dimensional Weyl multiplet after a Kaluza--Klein truncation on $T^2$. Such a truncation means that all the fields are taken to be independent of the $T^2$ coordinates. The dictionary between the fields is first established at the linear level by considering the transformations of the fields under the various bosonic symmetries. In a second subsection, we provide the complete non-linear relations between the four and six-dimensional fields by matching their off-shell suspersymmetry transformations.

\subsection{Reduction of six-dimensional Weyl multiplet}
	
The six-dimensional $N=(2,0)$ Weyl multiplet is based on the gauging of the superconformal algebra $OSp(8^*|4)$. It is an off-shell multiplet whose detailed construction can be found in \cite{Bergshoeff:1999db}. Similar to the four-dimensional $N=4$ Weyl multiplet, it contains the gauge fields that are associated with the gauge symmetries as well as various auxiliary fields. As usual, some of the gauge fields are dependent, and can be expressed in terms of the other fields through supercovariant curvature constraints. The independent fields of the $N=(2,0)$ Weyl multiplet and their properties are summarized in Table~\ref{table:6dWeyl}. In six dimensions, the R-symmetry group is $USp(4)$ such that the $i,j$ indices still run from 1 to 4. All the bosonic fields are real, while all the spinors are $USp(4)$ symplectic Majorana spinors that moreover satisfy definite properties with respect to the six-dimensional chirality matrix $\Gamma_*$. We use $M,N,\ldots$ and $A,B,\ldots$ for the six-dimensional spacetime and tangent space indices, respectively. The infinitesimal transformations of the various fields under $Q$-supersymmetry (with a parameter $\epsS^i$) and $S$-supersymmetry (with a parameter $\etaS^i$), are given by\footnote{Our convention differs from that of \cite{Bergshoeff:1999db} by $\eS_{M}{}^{A}\to-\eS_{M}{}^{A}$.}
\begin{subequations}
\label{6dsusy}
\begin{align}
\d \eS_M{}^A =&\,-\tfrac{1}{2}\bar\epsS^i\,\Gamma^A\psiS_{M\,i}\,,\\
\d \boldsymbol{\widehat b}_M=&\, -\tfrac12\bar\epsS^i\,\boldsymbol{\widehat\phi} _{M\,i}+\frac12 \bar\etaS^i \psiS_{M\,i}\,,\label{eq:6DSUSYb}\\ 
\d \psiS_M^i =&\, \DS_M\epsS^i - \tfrac{1}{24}
\TS^{ij}_{ABC}\Gamma^{ABC}\Gamma_M\epsS_j - \Gamma_M \etaS^i
\,,\\
\d \VS_M^{ij} =&\, -4{\bar\epsS}^{(i} \boldsymbol{\widehat \phi}_M^{j)} +
{\tfrac{4}{15}}{\bar\epsS}_k\Gamma_M\chiS^{(i,j)k}-4{\bar\etaS}^{(i}\psiS_M^{j)}\,,\\
\d \TS_{ABC}^{ij}=&\, {\tfrac{1}{8}}{\bar\epsS}^{[i}
\Gamma^{DE}\Gamma_{ABC}\,{ \boldsymbol{ \widehat R}}(Q)^{j]}_{DE}
- {\tfrac{1}{15}} {\bar\epsS}^k\Gamma_{ABC}\,\chiS_k{}^{ij} - \text{$\Omega$-trace}\,,\\
\d \chiS_k{}^{ij} =&\,{-\tfrac{5}{32}}
\boldsymbol \DS_M \TS^{ij}_{ABC}\,\Gamma^{ABC}\Gamma^M\epsS_k -\tfrac{15}{16}
\Gamma^{MN} \boldsymbol{\widehat R}(V)_{MN\, k}{}^{[i} \,\epsS^{j]} -\tfrac{1}{4}\DS^{ij}{}_{kl}\,\epsS^l\nn\\*
&\quad +\tfrac58 \TS_{ABC}^{ij} \Gamma^{ABC} \etaS_k - \text{$\Omega$-trace}-\text{$\delta$-trace}\,,\\
\d \DS^{ij,kl} =&\, 2
{\bar\epsS}^{[i} \boldsymbol{ \widehat{ \slashed{D}} }\chiS^{j],kl}
+4 \bar\etaS^{[i}\chiS^{j],kl}
+ (ij\leftrightarrow kl ) -\text{$\Omega$-traces}\,,
\end{align}
\end{subequations}
where the bar on the supersymmetry parameters denotes the Majorana conjugate (see subsection~\ref{sec:spinorred} for the definition). The field $\widehat{\boldsymbol{\phi}}_M^i$ that appears in \eqref{eq:6DSUSYb} is the gauge field associated with the six-dimensional $S$-supersymmetry. It is a dependent field which is determined through the supercovariant curvature constraint $\Gamma^M  \boldsymbol{\widehat R}(Q)_{MN}^i=0$, and whose explicit expression can be found in~\cite{Bergshoeff:1999db}. The fully supercovariant derivative $\DS_M$ generally includes the connections
for local Lorentz, $USp(4)$ R-symmetry, dilatation, special conformal transformations, and $Q$- and $S$-supersymmetry

As mentioned before, the goal of this section is to study the dimensional reduction of the six-dimensional $N=(2,0)$ Weyl multiplet on the two-dimensional torus.
	\begin{table}[ht!]
 \begin{center}
   \scalebox{1.00}{
 \begin{tabular}{ c | c c c c }
 \hline 
 \hline
  Field & Description & Restrictions & $\mathrm{USp}(4)$ & $w$ \\
 \hline
 $\eS_M{}^A$ & sechsbein & & $\mathbf 1$ & $-1$\\[1mm]
 $\psiS_M^i$ & gravitini & $\Gamma_* \psiS_M^i = \psiS_M^i$ & $\mathbf 4$ & $-\frac{1}{2}$\\[1mm]
 $\VS_M^{ij}$ & $USp(4)$ gauge field & $\VS_M^{ij} = \VS_M^{ji}$ & $\mathbf{10}$ & $0$ \\[1mm]
 $\TS_{ABC}^{ij}$ & bosonic matter field &  $\TS_{ABC}^{ij} = - \TS_{ABC}^{ji}$, $\Omega_{ij}\TS_{ABC}^{ij} =0$ & $\mathbf 5$ & $1$\\[1mm]
 & & anti-self-dual \eqref{eq:asd}& & \\[1mm]
$\chiS_k{}^{ij}$ & fermionic matter field & $\chiS_k{}^{ij} = - \chiS_k{}^{ji}$, $\Omega_{ij} \chiS_k{}^{ij} = 0$ & $\mathbf{16}$ & $\frac{3}{2}$ \\[1mm]
& & $\chiS_k{}^{ik} = 0$, $\Gamma_* \chiS_k{}^{ij} = \chiS_k{}^{ij}$ & & \\[1mm]
$\DS^{ij,kl}$ & bosonic matter field & $\DS^{ij,kl} = - \DS^{ji,kl} = -\DS^{ij,lk} = \DS^{kl,ij}$ & $\mathbf{14}$ & $2$\\[1mm]
& & $\Omega_{ij}\DS^{ij,kl} = \Omega_{kl}\DS^{kl,ij} = \Omega_{ik}\Omega{jl}\DS^{kl,ij} = 0$ & & \\[1mm]
$\widehat{\boldsymbol{b}}_M$ & dilatation gauge field & & $\mathbf{1}$ & 0\\[1mm]
\hline
 \hline
\end{tabular}}
\caption{\textit{Fields of the (2,0) Weyl multiplet in six dimensions. $w$ denotes the Weyl weight while $\Gamma_{*}$ denotes the chirality gamma matrix in six dimensions. We have suppressed the dependent gauge fields that are expressed in terms of the fields above through conventional curvature constraints.}}
\label{table:6dWeyl}
\end{center}
\end{table}

{\bf Note on notation:}
We use bold hatted symbols for the fields of the six-dimensional Weyl multiplet and bold symbols for the four-dimensional fields arising after reduction. The latter will ultimately be related to the fields of the four-dimensional dilaton Weyl multiplet constructed in Section~\ref{sec:construction}, for which we will use a standard font. Furthermore, in six dimensions, the $USp(4)$ indices are raised and lowered using the invariant tensor $\Omega_{ij}$. We have for instance,  
\begin{align}
\psiS_M^{i}=\Omega^{ij}\psiS_{M\, j}
\quad \text{and} \quad
\psiS_{M\, i}=\psiS_M^{j}\Omega_{ji}\,.\label{eq:6Dindices}
\end{align}
Note that in six dimensions we are not using a chiral notation for the fields (unlike in four dimensions), so that the raising and lowering of indices in this way should not cause confusion. The reader should therefore keep in mind that the relations \eqref{eq:6Dindices} are not applicable in four dimensions (\textit{i.e.} for fields and parameters without a hat).

\subsubsection{Reduction of sechsbein}\label{subsec:sechsred}
	
Let us first consider the sechsbein $\eS_M{}^A$ which plays a central role in the dimensional reduction on $T^2$. It carries 15 off-shell degrees of freedom, and gives rise, after reduction to four dimensions, to three scalar fields, two Kaluza--Klein abelian vector fields, and a vierbein which respectively carry 3, (3+3), and 5 off-shell degrees of freedom. Among those scalar fields, the one parametrizing the overall size of internal torus will naturally be related to the dilaton field of the dilaton Weyl multiplet, while the other two will be related to the coset scalars. 

As is usual in Kaluza--Klein reduction, we first partially use the six-dimensional local Lorentz invariance to fix the sechsbein into an upper-triangular form. This means that we have
    \begin{align}
\label{eq:6Lfix}
\eS_\mu{}^a = \eR_\mu{}^{a}, \quad \eS_\alphatwo{}^a = 0, \quad \eS_\mu{}^\mathsf a = \eR_\alphatwo{}^\mathsf a\AR^\alphatwo_\mu, \quad \eS_\alphatwo{}^\mathsf{a} = \eR_\alphatwo{}^\mathsf{a}\,,
\end{align}
where $\AR^\alphatwo_\mu $ denotes the two Kaluza--Klein gauge fields. Here, $A \equiv \{a=\underline{0},\underline{1},\underline{2},\underline{3}, \mathsf a=\underline{4},\underline{5}\}$ are tangent space indices, while $\mu=\{0,1,2,3\}$ and $\alphatwo=\{4,5\}$ denote the curved indices for the four-dimensional spacetime and the internal torus, respectively.

The six-dimensional diffeomorphisms compatible with the isometries of the torus give rise to four-dimensional diffeomorphisms, $U(1)^2$ gauge transformations associated with the Kaluza--Klein gauge fields, and a rigid $GL(2,\mathbb{R})$ symmetry acting on the internal curved index of $\eR_\alphatwo{}^{\mathsf{a}}$ and $\AR^\alphatwo_\mu $. Note that the third equation in \eqref{eq:6Lfix} disentangles four-dimensional diffeomorphisms from $U(1)^2$ gauge transformations.

We now wish to establish the relations between the fields in \eqref{eq:6Lfix} and those of the four-dimensional dilaton Weyl multiplet. The first natural identification is $\eR_{\mu}{}^a = e_\mu{}^a$. To identify the other components of~\eqref{eq:6Lfix} with the dilaton Weyl fields, we introduce
	\begin{align}
	 \sR =\det(\eR_{\alphatwo}{}^{\mathsf{a}}) \,,\quad
	 \LR=\frac{1}{\sqrt{\sR}}\begin{pmatrix} \eR_{4}{}^{\underline{4}} & \eR_{5}{}^{\underline{4}}\\ \eR_{4}{}^{\underline{5}} & \eR_{5}{}^{\underline{5}} \end{pmatrix} \,, \quad
	 \AR_{\mu}=\begin{pmatrix} \AR_{\mu}^{4}\\\AR_{\mu}^{5} \end{pmatrix}\,,\label{eq:sechsbeincomp}
	\end{align}
which transform as
\begin{align}\label{SL2rtransf}
    \sR \rightarrow \frac{1}{\Lambda}\sR \,,\quad 
    \LR \rightarrow \LR\tilde{\Lambda}^{-1} \,, \quad
    \AR_{\mu} \rightarrow \Lambda^{1/2} \tilde{\Lambda}\AR_{\mu}
\end{align}
under the rigid $GL(2,\mathbb{R})$ transformations. Here, $\Lambda\in \mathbb{R}^+$ parametrizes the rigid scale transformation, while $\tilde{\Lambda}\in SL(2,\mathbb{R})$ is a two by two matrix. Upon inspecting the supersymmetry transformations of the four-dimensional Weyl multiplet (\ref{dil_weyl_trans}), one can check that they remain invariant under a rigid rescaling of the fields $\rhoF$, $\mathcal{\AF}_{\mu;\alpha}$ and $\psiF_{i}$ with the same weight. Based on this observation, we identify
\begin{align}
\label{eq:rhoI}
    \rhoF={4}\sR^{-1/2}\,,
\end{align}
which will be consistent with our identification (\ref{dil_weyl_gauge_id}) of $\mathcal{\AF}_{\mu;\alpha}$ with the Kaluza--Klein gauge fields $\AR_{\mu}^{\alphatwo}$. 

The dilaton Weyl multiplet contains the scalars $\phi_\alpha$ that parametrize $SU(1,1)/U(1)$ while those appearing in the torus reduction, \textit{i.e.} $\eR_\alphatwo{}^{\mathsf{a}}$, naturally parametrize $SL(2,\mathbb{R})/SO(2)$.
The groups $SL(2,\mathbb{R})$ and $SU(1,1)$ are isomorphic. For any $\tilde\Lambda\in SL(2,\mathbb{R})$, one can find $g\in SU(1,1)$ as $g=\Cayley\tilde\Lambda \Cayley^{-1}$, where $\Cayley$ is the Cayley matrix that we choose as
\begin{align}
    \Cayley=\frac{1}{\sqrt{2}}\begin{pmatrix} 1& i\\ -1& i\end{pmatrix}\,.\label{eq:Cayleymatrix}
\end{align}
Using this Cayley transformation, we can construct an $SU(1,1)$ matrix from the $SL(2,\mathbb{R})$ matrix $\LR^{-1}$,
\begin{equation}
U=\Cayley \LR^{-1} \Cayley^{-1}\,,\label{eq:fromLtoU}
\end{equation}
which we identify with the $SU(1,1)$ matrix $U$ parametrized by the coset scalars $\phi_\alpha$ of the dilaton Weyl multiplet (\ref{Udef1}). With this identification, the action of $SL(2,\mathbb{R})$ on $\LR$ in~\eqref{SL2rtransf} correctly reproduces the action of $SU(1,1)$ on $U$ in~\eqref{coset_transf}.\footnote{We have to use $\LR^{-1}$ since the global symmetry in~\eqref{coset_transf} acts from the left on $U$ and from the right on $\LR$ in~\eqref{SL2rtransf}.}
As a result, the scalars $\phi_\alpha$ are identified with the following internal components of the sechsbein
\begin{align}\label{coset_identif}
    \phiF_1&=\frac{1}{2\sqrt{\sR}}\left(\eR_{5}{}^{\underline{5}}+\eR_{4}{}^{\underline{4}}+i\eR_{5}{}^{\underline{4}}-i\eR_{4}{}^{\underline{5}}\right)\ \,, \quad\quad
    \phiF_2=\frac{1}{2\sqrt{\sR}}\left(-\eR_{5}{}^{\underline{5}}+\eR_{4}{}^{\underline{4}}-i\eR_{5}{}^{\underline{4}}-i\eR_{4}{}^{\underline{5}}\right)\,.
\end{align}

Consider now the four-dimensional $SU(1,1)$ covariant complex gauge fields $\mathcal{\AF}_{\mu;\alpha}$ of the four-dimensional dilaton Weyl multiplet, which transform under $SU(1,1)$ as in (\ref{GaugeFieldUtransf}). Since we know from (\ref{SL2rtransf}) how the rigid $SL(2,\mathbb{R})$ transformation $\tilde{\Lambda}$ acts on $\AR_{\mu}^\alphatwo$, and how the rigid $SU(1,1)$ transformation $g$ is related to $\tilde{\Lambda}$, we can relate the gauge fields $\mathcal{\AF}_{\mu;\alpha}$ of the dilaton Weyl multiplet with the Kaluza--Klein gauge fields $\AR_{\mu}^\alphatwo$ as
\begin{align}\label{dil_weyl_gauge_id}
    \begin{pmatrix} {\mathcal{A}}_{\mu;1} \\{\mathcal{A}}_{\mu; 2}\end{pmatrix}=\Cayley \begin{pmatrix} \AR_{\mu}^{4}\\\AR_{\mu}^{5}\end{pmatrix}\,,
\end{align}
where $\mathcal C$ is given by \eqref{eq:Cayleymatrix}.

Let us now finally comment on the four-dimensional local symmetries inherited from the six-dimensional Lorentz symmmetry. The transformation parameter $\betaS^{AB}$ decomposes into $(\betaR^{ab}, \betaR^{a \mathsf a},  \betaR^{\underline{5}\underline{4}}\equiv\betaR)$. Imposing the gauge fixing condition $\eS_{\alphatwo}{}^{a}=0$ 
breaks the six-dimensional Lorentz symmetry to the four-dimensional Lorentz symmetry (with parameter $\betaR^{ab}$) and a local $SO(2)$ symmetry (with parameter $\betaR$). However, the condition $\eS_{\alphatwo}{}^{a}=0$ is not preserved by $Q$-supersymmetry and as a result we have to redefine supersymmetry by a compensating six-dimensional Lorentz transformation. The new supersymmetry transformation then takes the form
\begin{align}
		 \delta_Q (\epsS) \coloneqq  \delta_Q^{old} (\epsS) +  \delta_L(\betaR^{a}{}_{\mathsf{a}}(\epsS))\;,
	\end{align}
where the non-vanishing component of the compensating parameter reads
	\begin{align}
\betaR^{a}{}_{\mathsf{a}}=-\frac{1}{2}\bar{\epsilonS}^{i}\Gamma^{a}\psiS_{\alphatwo i} \eS_{\mathsf{a}}{}^{\alphatwo}.
 \label{eq:Lorcomp}
\end{align}
The local $SO(2)$ transformation acts on the $SL(2,\mathbb{R})$ scalar matrix $\LR^{-1}$ defined in \eqref{eq:sechsbeincomp} as
\begin{align}
    \LR^{-1}  \rightarrow \LR^{-1} R 
    \quad\text{with}\quad
    R=\begin{pmatrix}
        \cos\betaR & \sin\betaR \\ -\sin \betaR & \cos\betaR
    \end{pmatrix}\,.
\end{align}
Under a Cayley transformation, this translates into the local $U(1)$ transformation that acts on the $SU(1,1)$ scalar matrix $U$ (\ref{Udef1}) from the right, i.e. as
\begin{align}
    U\rightarrow U \Omega
\quad\text{where}\quad
    \Omega=\Cayley R \,\Cayley^{-1}=\begin{pmatrix}
        e^{-i\betaR} & 0\\ 0& e^{i\betaR}
    \end{pmatrix}\,.
\end{align}
Together with \eqref{eq:fromLtoU}, this illustrates how the Cayley transformation relates the $SL(2,\mathbb{R})/SO(2)$ coset structure underlying the scalar sector of the sechsbein reduction and the $SU(1,1)/U(1)$ coset space encountered in~\eqref{coset_transf}, where we  identify $\betaR=\beta$. It can also be shown that the dependent $U(1)$ gauge field $\aaF_{\mu}$ in four dimensions directly descends from the internal component $\omegaS_{\mu}{}^{\underline{4}\underline{5}}$ of the six-dimensional spin connection.

\subsubsection{Reduction of other bosonic fields}\label{bos_oth}

Let us now consider the dimensional reduction of the $USp(4)$ gauge field $\VS_M^{ij}$ of the six-dimensional Weyl multiplet. We use the standard Kaluza--Klein decomposition, 
	\begin{align}
 \label{eq:Vredef}
		\VS_{M}^{ij}\,dx^M=\VR_\mu^{ij}\,dx^\mu + \VR_\alphatwo^{ij}\,\left(dy^\alphatwo + \AR^\alphatwo_\mu dx^\mu\right)\,.
	\end{align}	
The fields  $\VR_{\mu}^{ij}$ and $\VR_m^{ij}$ respectively transform as vectors and scalars under four-dimensional diffeomorphisms, and the inclusion of the Kaluza--Klein gauge fields $\AR_\mu^\alphatwo$ in the decomposition ensures that $\VR_{\mu}^{ij}$ is invariant under the $U(1)^2$ gauge symmetry. We may also use the internal component of the sechsbein \eqref{eq:6Lfix} to define the following scalars
	\begin{align}
 \label{eq:Vtoflat}
	\VR_{\mathsf{a}}^{ij} \coloneqq \eR_{\mathsf{a}}{}^\alphatwo\, \VR_\alphatwo^{ij}\,.
	\end{align}	
They are invariant under the rigid $GL(2,\mathbb{R})$ symmetry, but transform  under the local $SO(2)\cong U(1)$ originating from the internal part of the Lorentz transformations in six dimensions discussed at the end of~Section \ref{subsec:sechsred}.  (\textit{i.e.} those with parameter $\betaR^{\underline{5}\underline{4}}\equiv\betaR$). Let us then define the following linear combination of scalar fields,
\begin{align}
\label{eq:E4}
    \ER^{ij} \coloneqq \VR_{\underline{4}}^{ij} + i\VR_{\underline{5}}^{ij}, \qquad  \ER_{ij}\coloneqq (\ER^{ij})^* = \VR_{\underline{4}}^{ij} - i\VR_{\underline{5}}^{ij}\,,
\end{align}
which respectively carry a weight $c=+1$ and $c=-1$ under the local $U(1)$. These weights motivate the identification of the complex scalar $\ER_{ij}$ at the linear level with the scalar $E_{ij}$ of the four-dimensional dilaton Weyl multiplet. On the other hand, the field $\VR_{\mu}^{ij}$ is naturally related at the linear level to the $USp(4)$ gauge field $V_\mu{}^{ij}$ of the dilaton Weyl multiplet.

The antisymmetric tensor $\TS_{ABC}^{ij}$ generally decomposes into the following four-dimensional fields: $\TR^{ij}_{a\,\mathsf{ab}}$, $\TR_{ab\,\mathsf{a}}^{ij}$, $\TR_{abc}^{ij}$.\footnote{There is no need for Kaluza--Klein redefinitions of the type~\eqref{eq:Vredef} since the field carries tangent space indices.} Recall however that the six-dimensional field is anti self-dual, 
\begin{align}
 \label{eq:asd}
	\TS_{ABC}^{ij} = -\frac{1}{6}\hat \epsilon_{ABCDEF}\,\TS^{DEF\,ij}\,.
	\end{align}
Thus, the resulting four-dimensional fields are not going to be independent. Once we dimensionally reduce this duality relation, the algebraically independent components are $\TR_{a\, \mathsf{a}\mathsf{b}}^{ij}$ and $\TR_{ab\,\mathsf{a}}^{ij}$, where the latter satisfies
	\begin{align}
	    \TR_{ab\,\mathsf{a}}^{ij} &= -\frac{1}{2}\hat \epsilon_{abcd\,\mathsf{ab}}\TR^{cd\mathsf{b}\,ij} \,,\\ 
	    \Rightarrow\quad  \TR_{ab\,\underline{4}} ^{ij}\pm i\TR_{ab\,\underline{5}}^{ij} &= \pm \frac{1}{2}i\, \hat{\epsilon}_{abcd \underline{4}\underline{5}} \left(\TR^{cd\,\underline{4}\,ij} \pm i\TR^{cd\,\underline{5}\,ij}\right)\,.\nn
	\end{align}
By identifying $\hat \epsilon_{abcd\,\underline{45}} = \epsilon_{abcd}$, we see that $(\TR_{ab\,\underline{4}}^{ij} \pm i\TR_{ab\,\underline{5}}^{ij})$ is (anti) self-dual in four dimensions. Hence we define the following reduced fields 
\begin{align}
\label{def:tabij}
 \TR_{ab}^{ij}\coloneqq\TR_{ab\,\underline{4}}^{ij} - i\TR_{ab\,\underline{5}}^{ij}\,,\qquad  \TR_{ab\,ij}\coloneqq(\TR_{ab}{}^{ij})^*=\TR_{ab\,\underline{4}}^{ij} + i\TR_{ab\,\underline{5}}^{ij}\,,
\end{align}
which will be related at the linear level to the fields $T_{ab}^{ij}$ and $T_{ab\,ij}$ of the dilaton Weyl multiplet. The field 
\begin{align}
\label{def:taij}
    \TR_a^{ij} \coloneqq \TR_{a\, \underline{45}}^{ij} \,,
\end{align}
on the other hand will be related at the linear level to $X_a{}^{ij}$. 

Finally, the dimensional reduction of the six-dimensional scalars $\DS^{ij,kl}$ simply leads to four-dimensional scalars which will be denoted by $\DR^{ij,kl}$. These fields will be related at the linear level to the dilaton Weyl fields $D^{ij}{}_{kl}$. 

\subsubsection{Reduction of spinors}
\label{sec:spinorred}
For the dimensional reduction of spinors we first relate the Clifford algebras in  six and four space-time dimensions.  We can write a representation of the six-dimensional Clifford algebra ($\Gamma$) in terms of the four-dimensional Clifford algebra ($\gamma$) as follows
	\begin{gather}
		\Gamma^a \equiv 
		\begin{pmatrix} 
			\gamma^a & 0\\ 
			0 & \gamma^a 
		\end{pmatrix} \,,\quad \quad
		\Gamma^{\underline{4}} \equiv 
		\begin{pmatrix} 
			0 & \gamma_*\\ 
			\gamma_* & 0 
		\end{pmatrix}\,, \quad \quad
		\Gamma^{\underline{5}} \equiv 
		\begin{pmatrix} 
			0 & -i\gamma_*\\ 
			i\gamma_* & 0 
		\end{pmatrix}\,, \quad \quad
		\Gamma_* \equiv 
		\begin{pmatrix} 
			\gamma_* & 0\\ 
			0 & -\gamma_* 
		\end{pmatrix}		\,,
	\end{gather}	
where we recall that all underlined indices are flat and define $\gamma_*=i \gamma^{\underline{0}}\gamma^{\underline{1}}\gamma^{\underline{2}}\gamma^{\underline{3}}$ so that $\gamma_*^2=+1$ as well as $\Gamma_*= \Gamma^{\underline{0}}\Gamma^{\underline{1}}\Gamma^{\underline{2}}\Gamma^{\underline{3}}\Gamma^{\underline{4}}\Gamma^{\underline{5}}$ with $\Gamma_*^2=+1$. In both four and six dimensions, the gamma matrices are hermitian except for the anti-hermitian time-like ones $\gamma^{\underline{0}}$ and $\Gamma^{\underline{0}}$. We use $(-+\ldots +)$ signature and a representation such that $\gamma_*$ is real.

The spinors in six dimensions transform in the fundamental representation of the R-symmetry group $USp(4)$ within the superconfomal group $OSp(8^*|4)$. 
Writing an elementary spinor as $\psiS{}^i$, its (Majorana) conjugate spinor is defined as usual by $\bar\psiS{}^i = (\psiS{}^i)^T C$  where $C$ is the six-dimensional charge conjugation matrix satisfying $(C \Gamma^A)^T = - C \Gamma^A$ in six dimensions. Our choice is
\begin{align}
C
= \begin{pmatrix}
 0 & -c \\
 c & 0 
\end{pmatrix}
\end{align}
in terms of the real, antisymmetric  four-dimensional charge conjugation matrix $c$ that satisfies $(c\gamma^a)^T = +c\gamma^a$.
The six-dimensional charge conjugation matrix satisfies $C^T=C$. 
The charge conjugate of a six-dimensional spinor is given by $(\psiS{}^i)^C = - i \Omega^{ij} \Gamma^{\underline{0}} C^{-1} (\psiS{}^j)^*$ and this intertwines the symplectic structure with the Clifford structure. Because of this intertwining, one can consistently impose the symplectic Majorana condition $\psiS{}^i =  (\psiS{}^i)^C$.
 In addition to the $USp(4)$ symplectic Majorana condition one can impose a chirality condition $\Gamma_* \psiS{}^i = \pm \psiS{}^i$ in six dimensions and we call the corresponding spinors left- (for $+$) or right-handed (for $-$). The total number of real components of a chiral symplectic Majorana spinor is then 16. 
 In Table~\ref{table:6dWeyl} we have indicated the chiralities of the spinor fields. Note also that the parameter $\epsS^i$ of six-dimensional $Q$-supersymmetry is left-handed while the parameter $\etaS^i$ of $S$-supersymmetry is right-handed.
 
In four dimensions, we have the usual notion of Majorana spinors with four real components for which the two chiral components with respect to the chirality matrix $\gamma_*$ are related by complex conjugation. 
 The reduction of a left-handed  symplectic Majorana spinor $\psiS{}^i_L$  in six dimensions to four dimensions is then achieved by the formula	
\begin{align}
\label{eq:spinred}
\psiS{}^i_L = 
	\begin{pmatrix}
	\psiR^i \\ \Omega^{ij} \psiR_j
	\end{pmatrix}\,.
\end{align}
Here, $\psiR^i$ and $\psiR_i$ denote the left-handed and right-handed  parts  of a four-dimensional Majorana spinor. By slight abuse of notation we use the upper and lower indices to keep track of the two chiral projections in four dimensions, i.e. there is a four-dimensional Majorana spinor $\eta^i$ with $\psiR^i= \frac12(1+\gamma_*) \eta^i$ and $\psiR_i= \frac12(1-\gamma_*) \eta^i$. We note that a similar procedure can be performed for right-handed spinors in six dimensions where the handedness of the components in the reduction~\eqref{eq:spinred} are interchanged.

For spinors with multiple $USp(4)$ indices, the construction generalizes. The case of relevance is the left-handed spinor $\chiS_i{}^{jk}$ that reduces as
\begin{align}\label{chi_red}
    \chiS_i{}^{jk} =  \begin{pmatrix}
        \chiR_{i}{}^{jk} \\
        \Omega_{li}\Omega^{jm}\Omega^{kn} \chiR^l{}_{mn}
    \end{pmatrix}\,.
\end{align}

A vector-spinor, such as the gravitini $\psiS_M^i$, can be reduced by simply following the prescription for vector and spinor reductions independently. This extends to other mixed objects as well. For instance, \begin{equation}\label{vec_spi_red}
    \psiS{}^i_\mu = 
	\begin{pmatrix}
	\psiR^i_\mu + \psiR^i_\alphatwo \AR^\alphatwo_\mu \\ \Omega^{ij} \left(\psiR_{\mu\,j} + \psiR_{\alphatwo\, j} \AR^\alphatwo_\mu\right) 
    \end{pmatrix}\,,\quad\quad
\psiS^i_\alphatwo
    = 
	\begin{pmatrix}
	\psiR^i_\alphatwo  \\ \Omega^{ij}  \psiR_{\alphatwo\, j} 
    \end{pmatrix}\,.
\end{equation} 
With these definitions the compensating transformation $\betaR^{a}{}_{\mathsf{a}}$ in~\eqref{eq:Lorcomp} takes the form
\begin{align}
\betaR^{a}{}_{\mathsf{a}} = \frac{1}{2}{\bar{\epsR}^i}\hspace*{1pt}\gamma^a\psiR_{\alphatwo\,i} \eR_{\mathsf{a}}{}^\alphatwo +\frac{1}{2}{\bar{\epsR}}_i\hspace*{1pt}\gamma^a\psiR^i_\alphatwo  \eR_{\mathsf{a}}{}^\alphatwo \;.
\end{align}
in terms of the reduced spinors.

The four-dimensional components of the reduced gravitini $\psiR_\mu^i$ will be related to the gravitini of the dilaton Weyl multiplet $\psi_\mu^i$. We arrange the internal components according to
\begin{align}
\label{eq:defsigtau}
    \sigmaR^i &\coloneqq \psiR_{\underline{4}}^i + i\psiR_{\underline{5}}^i\,,
    \quad
\tauR_i \coloneqq \psiR_{\underline{4}\,i} + i\psiR_{\underline{5}\,i} \,.
\end{align}
Here, as in~\eqref{eq:Vtoflat}, we have taken the tangent space components since they transform in the same way as the dilaton Weyl fields under $U(1)\cong SO(2)$ and are invariant under $SU(1,1)\cong SL(2,\mathbb{R})$. These fields will be related  with the dilaton Weyl fields $\psi^i$ and $\Lambda^i$.

\subsubsection{Summary of the reduction of the six-dimensional Weyl multiplet}

The various four-dimensional fields that result from the Kaluza--Klein reduction of the six-dimensional Weyl multiplet are summarized in Table~\ref{tab:redfields}.
Note that the dilatational gauge field $\widehat{\boldsymbol{b}}_M$ generally decomposes into the four-dimensional dilatational gauge field $\boldsymbol{b}_\mu$ and two scalars $\boldsymbol{b}_\alphatwo$. In the following, we set the latter to zero using a special conformal transformation along the internal torus. This implies that the supersymmetry transformations must be redefined by a field-dependent special conformal transformation in order to maintain this gauge choice. However, since such a transformations only acts on the field $\boldsymbol{b}_\alphatwo$ itself, it will not play any role.
	\begin{table}[t!]
		\centering \renewcommand{\arraystretch}{1.5}
		\begin{tabular}{c c c c c}
		\hline\hline
			\textbf{D = 6} & & \textbf{D = 4} && defined in\\
		\hline
			\multirow{3}{*}{Sechsbein, $ \eS_M{}^A$} &  \multirow{3}{*}{$\longmapsto$} & 
   $\eR_\mu{}^a$ && \eqref{eq:6Lfix}\\
		& & 
  $\AR_\mu^m$ && \eqref{eq:6Lfix}
		\\
		& & 
  $\eR_m{}^\mathsf a$ && \eqref{eq:6Lfix} \\
		\hline
			\multirow{2}{*}{Gravitini, 
   $\psiS{}^i_M$ }&  \multirow{2}{*}{$\longmapsto$} & 
   $\psiR^i_\mu$ &&\eqref{vec_spi_red}\\
		& & 
  $\sigmaR_i,\: \tauR_i$ && \eqref{eq:defsigtau}\\
		\hline
			\multirow{2}{*}{$USp(4)$ gauge field, 
   ${\VS}^{ij}_M$} &  \multirow{2}{*}{$\longmapsto$} &
   $\VR_\mu^{ij}$ && \eqref{eq:Vredef}\\
		& & 
  $\ER^{ij}$ && \eqref{eq:E4} \\
  \hline
            \multirow{1}{*}{Dilatational gauge field, $\widehat{\boldsymbol{b}}_M$} &
            \multirow{1}{*}{$\longmapsto$} &
            $\boldsymbol{b}_\mu$ && see current section
            \\
		\hline
			\multirow{2}{*}{${\TS}_{abc}^{ij}$} &  \multirow{2}{*}{$\longmapsto$}  & $\TR_{ab}^{ij}$&& \eqref{def:tabij}\\
		  && $\TR_a^{ij}$  &&\eqref{def:taij}\\
		\hline
			${\chiS}_k{}^{ij}$ & $\longmapsto$ & $\chiR_k{}^{ij}$&&\eqref{chi_red}\\
		\hline
			${\DS}^{ij,kl}$ & $\longmapsto$ & $\DR\indices{^{ij,kl}}$ && see Section~\ref{bos_oth}\\
		\hline\hline
		\end{tabular}
		\caption{\textit{Dimensional reduction of the six-dimensional (2,0) Weyl multiplet fields. As in Table~\ref{table:6dWeyl}, we have suppressed the dependent gauge fields. The component $\boldsymbol{b}_\alphatwo$ is not displayed since it is gauge fixed to zero.} }
		\label{tab:redfields}
	\end{table}

\subsection{Supersymmetry and non-linear field redefinitions}
\label{sec:dict}
The dimensionally reduced fields of the six-dimensional Weyl multiplet and the fields of the dilaton Weyl multiplet constructed in Section \ref{sec:construction} transform in the same representations under the bosonic symmetries, but they do not yet transform identically under supersymmetry. This is because the relations between these fields have so far only been established at the linear order, and that additional non-linear modifications of some of these relations are necessary to match the supersymmetry transformations.

In addition to these non-linear field redefinitons, we also have to allow for modifications of the non-linear supersymmetry transformations. 
We therefore start from the following Ansatz for the modification of the reduced supersymmetry transformations, 
\begin{equation}
	    \delta_Q(\epsR) \longrightarrow \delta_Q(\epsR) + \delta_{USp(4)}(\Lambda^{ij}(\epsR)) + \delta_{U(1)}(\lambda(\epsR)) + \delta_S (\eta^i(\epsR)) \,,
	    \label{eq:susymod}
	\end{equation}
where $\Lambda^{ij}(\epsR)$, $\lambda(\epsR)$ and $\eta^i(\epsR)$ are field-dependent gauge parameters that remain to be determined.
We will fix the above $Q$-supersymmetry modifications and the non-linear field redefinitions at the same time, by comparing the supersymmetry variations of the dilaton Weyl multiplet fields and of the reduced six-dimensional fields that are in the same representations of the bosonic symmetries. 

We start from the assumption that all the fields coming from the sechsbein have the exact identifications already deduced in Section~\ref{subsec:sechsred}. That  is,
\begin{align}
     e_\mu{}^a&=\eR_{\mu}{}^a\,,\quad
     U= \Cayley \LR^{-1} \Cayley^{-1}\,,\quad
     \rho = {4}\sR^{-1/2}\,,\quad
     \begin{pmatrix} {\mathcal{A}}_{\mu;1} \\{\mathcal{A}}_{\mu; 2}\end{pmatrix}=\Cayley \begin{pmatrix} \AR_{\mu}^{4}\\\AR_{\mu}^{5}\,,\end{pmatrix}\label{eq:fixedid}
\end{align}
are assumed not to receive non-linear modifications. Moreover, we choose $\epsilonF^i=\frac12 \epsR^i$ which leads to the exact gravitini  identification $\psi_\mu^i=\psiR_\mu^i$ from the supersymmetry transformations~\eqref{6dsusy} and~\eqref{dil_weyl_trans}. 
 This hinges on the fact that the variation of $\eR_\mu{}^a$ is insensitive to the modifications in~\eqref{eq:susymod}.
Because $\sR$ is also insensitive to the supersymmetry modifications in \eqref{eq:susymod},  we can use its variation to derive the relation ${4}\rho^{-1} \psi^i=\sigmaR^i$. 
The additional factor of $\rho$ is consistent with the Weyl weights $1/2$ 
for  $\sigmaR^i$ and $3/2$ for $\psi^i$.

We then start fixing the field-dependent gauge parameters of the $Q$-supersymmetry modifications in~\eqref{eq:susymod}. The first step is to consider fields that only transform under $U(1)$ and that are invariant under the other symmetries. The only reduced fields that satisfy these properties are the scalars $\eR_m{}^\mathsf{a}$, which have already been identified with the scalars $\phi_\alpha$ through the second relation in~\eqref{eq:fixedid}. Their supersymmetry variation therefore allows us to determine the field-dependent gauge parameters $\lambda(\epsR)$, and also yields the dictionary for the field $\LambdaF_i$. 

The next step is to consider the variation of the expression obtained for the field $\LambdaF_i$ itself, which is invariant under $S$-supersymmetry, but transforms under $USp(4)$. Comparing it with the supersymmetry variation~\eqref{eq:dwLambda} of $\Lambda_i$, which is expressed in terms of the dilaton Weyl fields, allows us to fix the field-dependent $USp(4)$ gauge parameter $\Lambda^{ij}(\epsR)$ as well as the dictionary for the fields $\TF_{ab}^{ij}$ and $\EF_{ij}$.

The final parameter to be determined is $\eta^i(\epsR)$ and at this point we can consider any field that transforms under $S$-supersymmetry. We choose  the gravitino  whose dictionary was already determined below~\eqref{eq:fixedid}. Comparing its supersymmetry variation~\eqref{eq:dwgrav} in dilaton Weyl variables to the one of $\psiR_\mu^i$ obtained from~\eqref{6dsusy} then yields $\eta^i(\epsR)$, and allows us to extract the dictionary for the field $\XF_a{}^{ij}$. 

This completes the derivation of all the field-dependent gauge parameters appearing in \eqref{eq:susymod}. We give below their expression in terms of the dilaton Weyl fields,
\begin{subequations}
    \begin{align}
        \Lambda^{ij}(\epsilonF) &= \: -\frac{{4}}{\rhoF} \left(\bar \epsilonF^{(i} \psiF^{j)} + \Omega^{im}\Omega^{jn} \bar \epsilonF_{(m}\psiF_{n)} \right)\,,\\
        \lambda(\epsilonF) &= \: \frac{{2}i}{\rhoF} \left(\Omega_{mn} \bar \epsilonF^{m} \psiF^n - \Omega^{mn} \bar \epsilonF_m \psiF_n \right)\,,\\
        \eta^i(\epsilonF) &= \: -\frac{{2}}{\rhoF} \bar \psiF^{(i}\LambdaF_j \Omega^{s)j} \epsilonF_s  + \frac 1 4 \left(\frac{{16}}{\rho^2} \bar \psiF^{(i} \gamma_a \psiF_m + \bar\LambdaF_j \gamma_a \LambdaF^l \Omega_{lm} \Omega^{j(i} \right)\Omega^{k)m}\Omega_{ks}\gamma^a \epsilonF^s \nonumber\\*
        &\quad + \frac 1 {16}\left( \frac{{16}}{\rho^2} \bar \psiF^m \gamma_a \psiF_m + \bar\LambdaF_m \gamma_a \LambdaF^m\right) \gamma^a \epsilonF^i - \frac{3}{{2}}\Omega^{is}\frac{\tilde{F}_{bc}}{\rho}\gamma^{bc}\epsilonF_s + X_a{}^i{}_k \gamma^a \epsilonF^k\,,
    \end{align}
\end{subequations}
where $\widetilde{F}_{ab}$ was defined in~\eqref{FFF}.

Now that the modifications in~\eqref{eq:susymod} have all been determined, we can derive the rest of the dictionary between the reduced fields and the dilaton Weyl fields by repeatedly applying $Q$-supersymmetry on the previously established dictionary relations. After some lengthy algebra, this procedure unambiguously leads to the following complete list of non-linear relations, where the reduced fields appear on the right and the dilaton Weyl fields on the left, 
\begin{subequations}
\label{eq:finalid}
	\begin{align}
 	\epsilon^i &\equiv \tfrac12\epsR^i\,,\\
    \eta^i &\equiv \etaR^i \,,\\
    e_\mu{}^a &\equiv \eR_\mu{} ^{a}\,,\\
    \psi_\mu^i &\equiv \psiR_\mu^i \,,\\
 \rhoF &\equiv {4}\sR^{-\frac{1}{2}}\,,\\
    \phiF_1 &\equiv \frac{1}{2\sqrt{\sR}}\left(\eR_{5}{}^{\underline{5}}+\eR_{4}{}^{\underline{4}}+i\eR_{5}{}^{\underline{4}}-i\eR_{4}{}^{\underline{5}}\right)\,,\\
    \phiF_2 &\equiv \frac{1}{2\sqrt{\sR}}\left(-\eR_{5}{}^{\underline{5}}+\eR_{4}{}^{\underline{4}}-i\eR_{5}{}^{\underline{4}}-i\eR_{4}{}^{\underline{5}}\right)\,,\\
    \frac{{4}\psiF^i}{\rhoF} &\equiv \sigmaR^i\,,\\
    \LambdaF^i &\equiv -\Omega^{ij}\tauR_j\,,\\
    \EF_{is} &\equiv -\ER_{is} - \bar\sigmaR_i\sigmaR_s - \Omega_{im}\Omega_{sn}\bar\sigmaR^{(m}\tauR^{n)}\,,\\
    \mathcal{\AF}_{\mu;1} &\equiv\frac{1}{\sqrt{2}}\left(\AR_{\mu}^{4}+i\AR_{\mu}^{5}\right), \quad \mathcal{\AF}_{\mu;2} = -\left(\mathcal {\AF}_{\mu;1}\right)^*\,,\\
    \TF_{bc}^{kl} &\equiv \frac{1}{2}\TR_{bc}^{kl} - \frac{1}{4}(\bar\sigmaR^{[k}\gamma_{bc}\tauR^{l]} - \text{$\Omega$-trace})\,,\\
    \omegaF\indices{_\mu^{bc}} &\equiv - \omegaR\indices{_\mu^{bc}}\,,\\
    \bbF_\mu &\equiv \bbR_\mu\,, 
    \\
    \aaF_\mu &\equiv -\omegaS_\mu{}^{\underline{45}} - \frac{1}{4}i\,\Omega_{mn}\bar\sigmaR^m\psiR_\mu^n + \frac{1}{4}i\,\Omega^{mn}\bar\sigmaR_m\psiR_{\mu\,n} +\frac{1}{8}i\,(\bar\sigmaR^k\gamma_\mu\sigmaR_k + \bar\tauR^k\gamma_\mu\tauR_k)\,,\\
    \VF_\mu{}^{ij} &\equiv \frac{1}{2} \VR_\mu{}^{ij} + \frac{1}{2}(\bar\sigmaR^{(i}\psiR_\mu^{j)} + \Omega^{ik}\Omega^{js}\bar\sigmaR_{(k}\psiR_{\mu\,s)}) +\frac{1}{4}\left(\bar\sigmaR^{(i}\gamma_\mu \sigmaR_m + \bar\tauR^{(i}\gamma_\mu \tauR_m \right)\Omega^{j)m} \,,\\  
    \XF_{a}{}^{ij} &\equiv -i\,\TR_{a}^{ij} +\frac{1}{4}\left(\bar\sigmaR^{[i}\gamma_a \sigmaR_m + \bar\tauR^{[i}\gamma_a \tauR_m\right)\Omega^{j]m} - \text{$\Omega$-trace}\,,\\
    \chiF_k{}^{ij} &\equiv \frac{4}{15}\chiR_k{}^{ij} - i\gamma^a \sigmaR_k \TR\indices{_a^{ij}} - \frac{1}{2}\sigmaR^{[i}\Omega^{j]s}\ER_{sk} + \frac{1}{8}\gamma^{ab}\sigmaR^{[i}\TR\indices{_{ab}^{j]l}}\Omega_{lk} - \frac{1}{2}\tauR^{[i}\ER^{j]s}\Omega_{sk}  \nn\\
    &\quad+  \frac{1}{2}\bar\sigmaR_s\sigmaR_k\sigmaR^{[i}\Omega^{j]s} - \frac{1}{32}\bar\sigmaR^i\gamma^{ab}\sigmaR^j \gamma_{ab}\tauR^m\Omega_{mk} + \frac{1}{16}\bar\sigmaR^m\gamma^{ab}\sigmaR^{[i} \gamma_{ab}\tauR^{j]}\Omega_{mk} + \frac{1}{4}\bar\sigmaR_m\tauR_{k}\tauR^{[i}\Omega^{j]m} \nn\\
    &\quad + \frac{3}{4}\bar\sigmaR_k\tauR_m \tauR^{[i}\Omega^{j]m}  - \text{$\Omega$-trace}-\text{$\delta$-trace}\,,\\
    \DF^{ij}{}_{rs} &\equiv \Big[-\frac{1}{15} \DR\indices{^{ij,\,kl}} - \TR^{ij}\cdot \TR^{kl} + \frac{4}{15} \Omega^{m[i}\bar \sigmaR^{j]}\chiR_m{}^{kl} + \frac{4}{15} \Omega^{im}\Omega^{jn}\bar \sigmaR_{t}\chiR^{[k}{}_{mn}\Omega^{l]t} \nn\\*
    &\quad- i\TR_a^{ij}\bar\sigmaR^{[k}\gamma^a\sigmaR_m\Omega^{l]m}  + \frac{1}{8}\TR_{ab}^{ij}\bar\sigmaR^k\gamma^{ab}\sigmaR^l -\frac{1}{8}{\TR}_{ab}^{ij}\bar\sigmaR_p\gamma^{ab}\sigmaR_q\Omega^{p[k}\Omega^{l]q} + \frac{1}{2}\Omega^{p[k}{\ER}^{l][i}\Omega^{j]q}\ER_{pq}\nn\\*
    &\quad +\frac{1}{2}\Omega^{p[k}\ER^{l][i}\Omega^{j]q}\bar\sigmaR_p\sigmaR_q  + \frac{1}{2}\Omega^{p[k}\bar\sigmaR^{l]} \sigmaR^{[i}\Omega^{j]q}\ER_{pq} + \frac{1}{2}\Omega^{p[k}\bar\sigmaR^{l]}\sigmaR^{[i}\Omega^{j]q}\bar\sigmaR_p\sigmaR_q \nn\\*
    &\quad -i \TR\indices{_a^{ij}}\bar\tauR^{[k}\gamma^a\tauR_m\Omega^{l]m}  + \frac{1}{4}\bar\tauR^{[i}\gamma^a\tauR_m\Omega^{j]m}\bar\tauR^{[k}\gamma_a\tauR_n\Omega^{l]n} + \frac 1 2 \bar\sigmaR^{i}\sigmaR^{[k}\bar\sigmaR^{l]}\tauR^{j} - \frac 1 2 \bar\sigmaR^{j}\sigmaR^{[k}\bar\sigmaR^{l]}\tauR^{i} \nn\\*
    &\quad - \Omega^{p[k}\Omega^{l]m}\Omega^{n[i}\Omega^{j]q}\bar\sigmaR_{(m}\tauR_{n)}\bar\sigmaR_p\sigmaR_q + \frac 1 2 \ER^{i[k}\bar\sigmaR^{l]}\tauR^{j}  - \frac 1 2 \ER^{j[k}\bar\sigmaR^{l]}\tauR^{i}  \\*
    &\quad - \Omega^{p[k}\Omega^{l]m}\Omega^{n[i}\Omega^{j]q}\bar\sigmaR_m\tauR_n \ER_{pq} + 2 \bar\sigmaR_{(m}\tauR_{n)} \Omega^{n[i} \bar\tauR^{j]}\sigmaR^{[k}\Omega^{l]m} + \left( \text{$ij$} \leftrightarrow \text{$kl$} \right) -\text{traces}\Big]\Omega_{kr}\Omega_{ls}\,.		\nn
\end{align}
\end{subequations}	
The right-hand side of the last equation is only antisymmetric in $[ij]$ and $[kl]$. Note also that the subtraction of the traces projects it onto the ${\bf 14}$ representation of $USp(4)$.

The relations~\eqref{eq:finalid} provide the complete non-linear dictionary between the fields of the six-dimensional $N=(2,0)$ Weyl multiplet reduced on $T^2$ and the fields of the four-dimensional $N=4$ dilaton Weyl multiplet. This is one of the main results of the paper. These relations are easily established at the linear level based on the bosonic transformation properties of the fields. The non-linear relations are more difficult to derive, and required here a careful analysis of the supersymmetry transformations.

\section{Discussion}

In this paper, we have discussed the torus reduction of $N=(2,0)$ conformal supergravity in six dimensions to $N=4$ conformal supergravity in four dimensions. Our analysis of the reduction was only carried out at the kinematical level. More precisely, we have shown that the reduction of the six-dimensional $N=(2,0)$ Weyl multiplet leads to a variant of the $N=4$ Weyl multiplet in four dimensions, which is known as the dilaton Weyl multiplet. We constructed this off-shell multiplet explicitly for the first time, and presented the complete non-linear dictionary between its component fields and the reduced six-dimensional fields. 

The $N=4$ dilaton Weyl multiplet can in principle be leveraged to provide new (partially) off-shell descriptions of $N=4$ Poincar{\'e} supergravity.  When using the standard Weyl multiplet,  it was shown in \cite{deRoo:1984zyh} that a system of $6+n$ vector multiplets coupled to conformal supergravity is in fact gauge equivalent to Poincar{\'e} supergravity coupled to $n$ vector multiplets. In this case, the scalars and spin-1/2 fields of the extra $6$ vector multiplets play the role of compensators for some of the superconformal local symmetries.  The resulting Poincar{\'e} Lagrangian is invariant under rigid $SO(6,n)$ transformations that act linearly on the $6+n$ vector fields (and non-linearly on the scalars). This is part of the larger rigid $SU(1,1)\times SO(6,n)$ duality symmetry of $N=4$ Poincar{\'e} supergravity, but in this frame the $SU(1,1)$ factor is only realized at the level of the field equations. It would be interesting to reconsider the conformal description of $N=4$ Poincar{\'e} supergravity, by using instead the dilaton Weyl multiplet as a starting point. Due to the presence of the dilaton which can be used to gauge fix the local dilatation symmetry, one expects that fewer compensating multiplets will be required. Preliminary investigations suggest that five compensating vector multiplets might be sufficient for obtaining pure Poincar\'e supergravity. Based on the fact that $SU(1,1)$ is realized on the doublet of vectors $\mathcal{A}_{\mu;\alpha}$ in the dilaton Weyl multiplet, one also expects that the resulting $N=4$ Poincar{\'e} Lagrangian will be invariant under $SU(1,1)$. In this way, it therefore seems that one can recover $N=4$ Poincar{\'e} supergravity in a different symplectic frame.

The $N=4$ dilaton Weyl multiplet could also potentially be used for the construction of new off-shell invariants of $N=4$ conformal supergravity in four dimensions. The most general class of $N=4$ conformal supergravity invariants that only involve the standard Weyl mulitplet fields was derived in \cite{Butter:2019edc}. Their leading term consists of the Weyl tensor squared multiplied by a holomorphic function of the coset scalars. Since these scalars have zero Weyl weight, these terms are clearly invariant under local dilatations. For invariants based on the dilaton Weyl multiplet, powers of the dilaton field could in principle appear in various terms to ensure that their Weyl weight vanishes. One might then wonder about the possibility of constructing off-shell invariants that involve more than four derivatives by using the dilaton Weyl multiplet. 
In particular, one expects that the dimensional reduction of the $N=(2,0)$ invariant in six dimensions \cite{Butter:2017jqu,Butter:2016qkx}, which starts with a contraction of three Weyl tensors, should lead upon torus reduction to a six-derivative invariant of $N=4$ conformal supergravity. More generally, it would be interesting to study the construction of conformal supergravity invariants based on the dilaton Weyl multiplet by using the procedure outlined in \cite{Butter:2019edc}. We leave this for future work.

\subsubsection*{Acknowledgements}
We are grateful to Daniel Butter for many insightful comments and discussions. 
We would also like to thank Subramanya Hegde, Soumya Adhikari and Madhu Mishra for useful discussions. SM's research was supported by a DST Inspire fellowship and a DAAD WISE fellowship.
The research of FC and AK was supported in part by the European Research Council (ERC) under the European Union’s Horizon 2020 research and innovation programme (grant agreement No 740209). The research of FC is funded by the Deutsche Forschungsgemeinschaft (DFG, German Research Foundation) – 521509185. During the finalisation of this work, AK's research was supported by the Munich Institute for Astro-, Particle and BioPhysics (MIAPbP) which is funded by the Deutsche Forschungsgemeinschaft (DFG, German Research Foundation) under Germany's Excellence Strategy - EXC-2094 - 390783311. The research of BS was supported in part by the core research grant (CRG/2018/002373) of the Science and Education Research Board (SERB), Government of India. 
BS would like to thank Max Planck Institute for Gravitational Physics Potsdam (Albert Einstein Institute), Institute of Mathematical Sciences Chennai, Chennai Mathematical Institute, Indian Institute of Science Education and Research Bhopal and Harish Chandra Research Institute Prayagraj for hospitality during the course of the work.

	\bibliography{references}
	\bibliographystyle{jhep}

\end{document}